\documentclass[12pt]{article}

\usepackage[english]{babel}

\usepackage[letterpaper,top=2.5cm,bottom=2.5cm,left=2.5cm,right=2.5cm,marginparwidth=1.75cm]{geometry}
\usepackage{float}
\usepackage{authblk}

\usepackage{comment}
\usepackage{amsmath}
\usepackage{amsthm}
\usepackage{amsfonts}
\usepackage{graphicx}
\usepackage{csquotes}
\usepackage{booktabs}
\usepackage{multirow}
\usepackage{array}
\usepackage[colorlinks=true, allcolors=blue]{hyperref}
\usepackage[round]{natbib}

\usepackage{setspace}
\title{Coarsening-adjusted estimators for finite-population indicators}%era "addressing bias" 
\author[1]{Gaia Bertarelli}
\author[2]{Aldo Gardini}
\affil[1]{Department of Economics, Ca' Foscari University of Venice, Italy}

\affil[2]{Department of Statistical Sciences ``Paolo Fortunati'', University of Bologna, Italy\\ e-mail: \texttt{aldo.gardini@unibo.it}}

\date{}

\begin{document}
\maketitle

\begin{abstract}\noindent
%Self-reported numerical variables in sample surveys are often affected by coarsening, as respondents report rounded or heaped values rather than the underlying quantity of interest. This paper develops a general framework for estimating finite-population indicators from coarsened survey data under complex sampling designs. The observed response is treated as a coarsened manifestation of a latent variable, with reporting regimes of increasing coarseness modelled jointly with the latent distribution. The fitted model is used to generate posterior predictive replicates of the latent sample values, conditionally on the observed responses and auxiliary information. Standard design-based estimators are then applied to these replicated samples, preserving the finite-population target while propagating the additional uncertainty induced by coarsening. The resulting variance decomposition separates design uncertainty from the uncertainty due to the unobserved latent responses. The method is applied to smoking data from the Italian PASSI surveillance system, focusing on means, quantiles, and the prevalence of heavy smoking. A simulation study assesses the performance of the proposed estimators under alternative coarsening mechanisms.\\
%VERSIONE 2 accettata:
Self-reported numerical variables in sample surveys are frequently subject to coarsening, as respondents tend to report rounded or heaped values rather than the exact underlying quantity. While the statistical literature offers various model-based solutions, official statistics and public health surveillance routinely require design-based inference on finite-population indicators under complex sampling designs. This paper introduces a general framework that bridges this gap by treating the observed response as a coarsened manifestation of a latent variable, modeling reporting regimes of increasing coarseness jointly with the latent distribution via a survey-weighted pseudo-likelihood approach. The fitted model is used to generate posterior predictive replicates of the latent values, to which standard design-based estimators are applied, formally propagating the additional uncertainty through an explicit variance decomposition. Simulation studies assess the finite-sample performance and robustness of the proposed method under various misspecification and sampling scenarios. Finally, its practical relevance is demonstrated through an application to behavioural data from the Italian PASSI surveillance system, illustrating how coarsening affects different functionals, such as means, quantiles, and threshold-based prevalence indicators in heterogeneous ways.

\end{abstract}

\section{Introduction and motivating example}
%coarsening/heaping in survey sampling
%\begin{itemize}
%    \item \citep{drechsler2016beat}: stare attenti a questi. Fanno una cosa simile alla nostra idea però: framework di imputazione di Rubin, non propongono un framework troppo generale (target su reddito e basta) e dicono che la lognormale è adeguatissima a descrivere i redditi (...). Loro non hanno l'idea della variabile latente sotto, quindi l'idea della decomposizione della varianza. Secondo me noi possiamo: scrivere una procedura flessibile (vedi mistura), che va bene per tante cose. 
%    \item \citep{zinn2016statistical} self reported income, direi cosa diversa (puntano alla stima della distribuzione, non si occupano di sintesi).
%\end{itemize}

Sample surveys remain one of the main instruments through which national statistical institutes and public health agencies produce timely and policy-relevant indicators. Large-scale population surveys are routinely used to monitor socio-economic and health conditions, support evidence-based policy making, and provide representative estimates for target populations and domains \citep{un2014fundamental}. Public health surveillance systems, in particular, rely extensively on repeated or continuous surveys to monitor health behaviours, risk factors, and preventive practices in the population \citep{remington2010chronic, cdc_nhis}. In many domains, however, the variables of interest are not observed exactly. Respondents are often asked to report quantities such as income, expenditure, body weight, height, number of cigarettes smoked, alcohol consumption, minutes of physical activity, or time spent in sedentary behaviour \citep{tourangeau2000psychology, bound2001measurement, gross2016kernel}. Although these variables are conceptually continuous, or at least finely measured, respondents are often required to provide integer-valued answers, and their responses frequently display strong concentrations at a limited set of preferred values. These phenomena are usually referred to as rounding and heaping at preferred digits, respectively. Both can be viewed as instances of the more general concept of coarsening, which occurs when respondents report a simplified version of the underlying latent quantity rather than the quantity itself \citep{heitjan1991ignorability, heitjan1994ignorability, wang2008modeling}.

Coarsening is one of the main challenges affecting self-reported health survey data, together with other non-sampling errors such as recall error, social desirability bias, and the cognitive effort required to retrieve and summarize habitual behaviours \citep{eurostat_ehis}. 
The statistical literature recognizes coarsening as a general form of incomplete observation. \citet{heitjan1991ignorability} introduced a general framework for coarse data, treating rounding, heaping, censoring, grouping, and missingness as special cases, while \citet{heitjan1994ignorability} further formalized ignorability conditions for incomplete-data models by explicitly considering the observed degree of coarseness as part of the data structure. Early work by \citet{heitjan1990inference} also showed how multiple imputation can be used to draw inference from heaped responses when the exact latent value is not observed. A related perspective is provided by \citet{manski2010rounding}, who studied rounding in survey responses as a problem of partial information and identification. In the specific case of heaped numerical responses, several contributions have proposed models that jointly describe the latent value and the reporting mechanism. For instance, \citet{wang2008modeling} modelled heaping in self-reported cigarette counts, showing that respondents tend to report rounded counts at multiples of 5, 10, or 20, and that this reporting error can bias parameters of interest such as mean cigarette consumption. Related work has addressed heaping in self-reported income data. \citet{drechsler2016beat} proposed an imputation strategy for rounded income values based on the estimation of rounding probabilities and the re-imputation of reported values, whereas \citet{zinn2016heaping} focused on recovering the underlying income distribution from heaped responses. More recently, \citet{gardini2026coarsened} extended this perspective to small area estimation, proposing a Bayesian unit-level model for coarsened self-reported smoking data.

When self-reported survey data are used to construct official indicators, the consequences of coarsening are not limited to model parameters. Coarsening can affect estimates of population quantities such as means, quantiles, prevalence rates, and threshold-based indicators, which are routinely used to monitor public health targets, compare geographical areas, allocate resources, and evaluate prevention policies. Ignoring the reporting process may therefore lead to distorted evidence for policy decisions. Despite the progress made in the modelling literature, two aspects remain especially relevant for official statistics and public health surveillance. First, many applications require inference on finite-population quantities under complex sampling designs, rather than inference only on superpopulation model parameters. Second, the same coarsening mechanism may affect different functionals in different ways. A mean may be relatively stable under rounding, whereas a quantile or a threshold-based prevalence indicator may be highly sensitive to mass artificially accumulated at policy-relevant cut-points. This issue becomes particularly important when survey indicators are used to classify population subgroups, compare regions, or monitor targets defined through thresholds.

The motivating application of this paper comes from the Progressi delle Aziende Sanitarie per la Salute in Italia survey, known as PASSI, the Italian behavioural risk factor surveillance system coordinated by the Istituto Superiore di Sanità \citep{iss_2023_protocollo_passi, iss_passi_temi}. PASSI monitors health-related behaviours and preventive practices in the adult population aged 18--69 years, with the explicit aim of supporting prevention activities at the local level and evaluating public health objectives set out in national and regional prevention plans. The system is based on continuous sample surveys and collects information on lifestyles and risk factors related to chronic diseases. Official documentation reports high territorial coverage, with participating local health units covering more than 90\% of the Italian resident population and response rates above 85\%. These features make PASSI a central source for public health surveillance in Italy, but also a setting in which self-reported coarsening may directly affect indicators used for policy monitoring. Several PASSI variables are naturally exposed to heaping. Counts of cigarettes smoked per day tend to concentrate at multiples of 5 and 10; minutes of physical activity tend to concentrate at multiples of 10, 30, or 60; and anthropometric measures such as weight and height may also be rounded before being used to compute body mass index.

\begin{figure}
    \centering
\includegraphics[width=.9\linewidth]{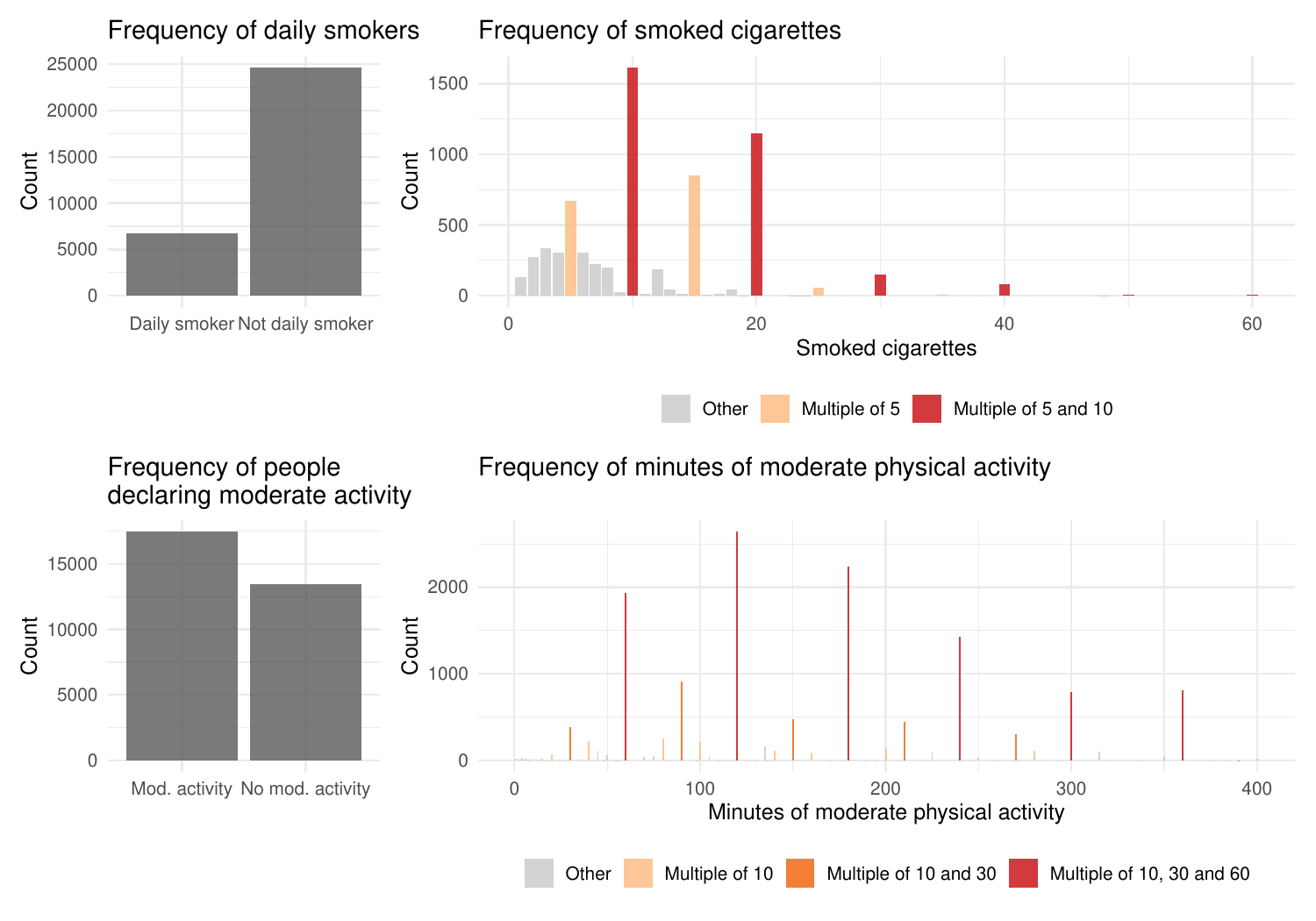}
    \caption{Empirical distributions of two self-reported PASSI variables: daily smoking status and number of cigarettes smoked per day among daily smokers; moderate physical activity status and reported minutes of moderate physical activity among active respondents. Colours distinguish increasingly coarse reported values.}
    \label{fig:esplorativa}
\end{figure}

Figure~\ref{fig:esplorativa} illustrates two motivating applications considered in this paper. The first concerns smoking behaviour, measured through daily smoking status and the number of cigarettes smoked per day among daily smokers. The second concerns moderate physical activity, measured through activity status and the reported duration in minutes among active respondents. In both cases, the positive part of the distribution displays clear heaping at preferred values. Cigarette counts concentrate at multiples of 5 and 10, with pronounced spikes at values such as 10 and 20 cigarettes per day, while reported minutes of moderate physical activity concentrate at multiples of 10, 30, and 60.

This coarsening is particularly relevant for threshold-based public health indicators. For smoking, a widely used indicator is the proportion of heavy smokers (HS), commonly defined using a threshold of 20 cigarettes per day \citep{eurostat_tobacco_2022}. For physical activity, indicators based on weekly duration often rely on policy-relevant cut-points, such as 120-150 minutes of moderate activity \citep{who2020physicalactivity}. In a continuous latent population, the probability of observing exactly the threshold would generally be negligible. In self-reported data, however, these thresholds may coincide with strong heaping points. As a result, estimated prevalences may depend non-negligibly on whether individuals reporting exactly the threshold value are classified as above or below it, showing how the measurement process can directly affect policy-relevant quantities.
The same issue is not limited to threshold-based indicators. When reported values are concentrated at preferred numbers, quantiles and other distributional summaries may become locked at heaping points, so that their estimates and confidence intervals partly reflect the discreteness of the reported distribution rather than the uncertainty about the underlying latent behaviour. Means may be less affected, because positive and negative rounding errors can partly compensate. This heterogeneity across estimands is central to our analysis: the impact of coarsening cannot be assessed only in terms of average bias, but must be evaluated with respect to the specific indicators used by statistical offices and public health authorities.

These considerations motivate the need for methods that account for coarsening while retaining the inferential targets and estimators used in survey practice. A closely related contribution is \citet{drechsler2016beat}, who study the effect of rounded income values on distributional indicators such as poverty rates and propose a multiple-imputation strategy to recover plausible unrounded values before analysis. Their approach is close in spirit to ours, as it replaces rounded responses with plausible unrounded values before estimating the quantities of interest. We extend this perspective by embedding the correction explicitly in a general design-based framework for finite-population indicators under complex sampling designs.

Building on the coarse-data inferential framework \citep{heitjan1991ignorability, heitjan1994ignorability}, we treat the observed response as a coarsened manifestation of an underlying latent variable. The coarsening mechanism is represented through reporting regimes of increasing coarseness, whose probabilities may depend on the latent value itself, allowing for non-ignorable reporting behaviour. The framework is general in the sense that different latent models, coarsening mechanisms, and survey estimators can be combined according to the structure of the variable and the target indicator.
The fitted model is used to generate posterior predictive replicates of the latent sample values, conditionally on the observed responses and auxiliary information. Standard survey estimators are then applied to these replicated samples, preserving a finite-population target while propagating the uncertainty due to coarsening. The resulting variance decomposition separates the usual design component from the additional uncertainty induced by the unobserved latent values.

The proposed strategy is assessed through a simulation study and through two applications to PASSI data, concerning cigarette consumption and moderate physical activity. These examples have different measurement structures and target indicators, and are used to illustrate how alternative modelling and estimation choices can be implemented within the same framework. The analysis shows how coarsening can affect means, quantiles, and threshold-based proportions in different ways.

The remainder of the paper is organized as follows. Section \ref{sec:procedure} formalizes coarsening in self-reported survey data and introduces the latent-variable representation adopted throughout the paper. Moreover, it develops the proposed coarsening-adjusted design-based estimator. Section \ref{sec:application} applies the method to PASSI smoking data, focusing on HS proportion, means, and quantiles of daily cigarette consumption. Section S3 of the supplementary material reports an additional application to moderate physical activity, illustrating how the framework can be adapted to a different self-reported behaviour. Section \ref{sec:sim} presents a simulation study designed to assess the performance of the proposed estimators under alternative coarsening mechanisms.  Section \ref{sec:conclusion} concludes the paper.

%\section{Coarsening in self-reported survey} {\color{red}{GB: serve questa sezione 2?}}
%\textcolor{magenta}{Forse no a questo punto, si richiama la figura nell'introduzione e buonanotte. }

\section{A design-based estimator adjusted for coarsening}\label{sec:procedure}

Let \(\mathcal U=\{1,\dots,N\}\) denote a finite population of size \(N\), from which a sample \(s\subseteq \mathcal U\) of size \(n\) is selected according to a possibly complex sampling design $p$. Let \(Y_i\) be the value of the study variable of interest for unit \(i\in\mathcal U\), and let \(\mathbf{Y}=\{Y_i\}_{i\in\mathcal U}\) denote the corresponding finite-population vector. We are interested in inference on a finite-population quantity \(\theta=\theta(\mathbf{Y})\). For the sampled units, let \(\mathbf{y}_s=\{Y_i\}_{i\in s}\) denote the sample values.

A design-based estimator of \(\theta\) is a function of the sampled observations and of the sampling design information, and can be written as a function of sample information
\[
\hat{\theta}=T(\mathbf{y}_s,\mathbf{w}_s),
\]
where \(\mathbf{w}_s=\{w_i\}_{i\in s}\) denotes the vector of sampling weights associated with the sampled units.
Besides the point estimator \(\hat{\theta}\), design-based inference also requires an assessment of its sampling variability. Let $
\mathbb{V}_p(\hat{\theta})
$
denote the design variance of \(\hat{\theta}\), and let
$
\widehat{V}_p(\hat{\theta})
$
be an estimator of this quantity. The choice of the estimator and its related variance estimator generally depends on both the target finite-population quantity and the features of the sampling design. Accordingly, different design-based estimators may be adopted, such as the Horvitz--Thompson estimator, the Hájek estimator, or other appropriate estimators \citep{sarndal1992}.

Confidence intervals are typically obtained by combining the point estimator
with the corresponding variance estimator. In the standard design-based
approach, one relies on a large-sample normal approximation to the sampling
distribution induced by the sampling design \citep{lumley2004survey}. Thus, the distribution of
\(\hat{\theta}\) under repeated sampling from the fixed finite population is
approximated by a normal distribution, leading to the approximate
\((1-\alpha)\)-level confidence interval
\(
\hat{\theta}
\pm
z_{1-\alpha/2}
\sqrt{\widehat{V}_p(\hat{\theta})}
\).
Alternative interval estimators, for instance based on replication methods or
bootstrap procedures for complex survey designs, may also be adopted. In
addition, specific interval construction methods may be preferable for particular
target quantities, such as proportions and other parameters with bounded
support, or for nonlinear functionals such as quantiles.

\subsection{Introducing coarsening}

Suppose now that the target variable \(Y_i\) is not directly observable and that, for each sampled unit \(i\in s\), only a manifest version \(Y_i^\star\) is observed. We assume that \(Y_i^\star\) is a coarsened version of the latent variable of interest \(Y_i\). Therefore, inference on the finite-population quantity \(\theta=\theta(\mathbf{Y})\) must rely on the observed sample values \(\mathbf{y}_s^\star=\{Y_i^\star\}_{i\in s}\), rather than on the unobserved latent values \(\mathbf{y}_s\). A naive estimator based on the observed data can thus be written as
\[
\hat{\theta}_\text{Naive}=T(\mathbf{y}_s^\star,\mathbf{w}_s).
\]

To formalize the coarsening mechanism, we assume that the manifest variable \(Y^\star\) is generated by two latent components: the latent true value \(Y\in\mathcal{Y}\) and a latent coarsening variable \(G\in\mathcal{G}\), which represents the reporting behaviour governing the observation process. Hence, the observed response \(Y^\star\) is assumed to arise from the joint action of \(Y\) and \(G\), that is,
\(
Y^\star = h_{Y^\star}(Y,G),
\)
for a suitable coarsening function \(h_{Y^\star}(\cdot,\cdot)\).

More specifically, let \(G\) take values in the finite set
\[
\mathcal G=\{c_1,\dots,c_j,\dots,c_J\},
\]
where each \(c_j\) represents a possible coarsening level. In particular, the
event \(G=c_j\) indicates that the respondent reports a coarsened value
\(Y^\star\) that is a multiple of \(c_j\). If \(0\in\mathcal G\), the value
\(G=0\) is used to denote exact reporting, so that non-coarsened records are
allowed in the sample. For each coarsening level \(g\in\mathcal G\), the observed value \(Y^\star\)
identifies the set of latent values \(Y\) that could have generated it. This
compatibility set is defined as
\[
I_g(Y^\star)
=
\begin{cases}
\{Y^\star\}, & g=0,\\[1mm]
\left[
Y^\star-\dfrac{g}{2},\,
Y^\star+\dfrac{g}{2}
\right), & g>0,
\end{cases}
\]
where \(g=0\) denotes exact reporting, while for \(g>0\) the level is admissible
only when \(Y^\star\) is a multiple of \(g\).

 In practice, the coarsening level is not directly observed and, since different pairs
\((Y,G)\) may lead to the same observed value \(Y^\star\), the manifest response
does not fully reveal the latent coarsening state.
Following \citet{heitjan1994ignorability}, we introduce the observed coarsening
information
\(
H=h_H(Y,G),
\)
where \(h_H\) is a coarsening-information map. Since the manifest response
\(Y^\star\) is itself generated from the pair \((Y,G)\), the realized value of
\(H\) can be characterized, in the present setting, as the set of coarsening
levels that are compatible with the observed response:
\[
H
=
\left\{
g\in\mathcal G:
Y^\star \text{ is admissible under coarsening level } g
\right\}.
\]
Thus, \(H\subseteq\mathcal G\) summarizes the recoverable information on the
latent coarsening state.

\subsection{A model for non-ignorable coarsening}\label{sec:model}

Within this framework, it is important to recover the latent response conditionally on the observed coarsened value. To this end, we formulate a statistical model capable of handling a non-ignorable coarsening mechanism, so as to predict the latent variable and construct a coarsening-adjusted (CA) design-based estimator of \(\theta\).

For each sampled unit \(i\in s\), let \(Y_i\) denote the latent response, \(G_i\) the latent coarsening level, \(Y_i^\star\) the observed coarsened response, and \(H_i\) the observable information on the coarsening process. Since both \(Y^\star\) and \(H\) are deterministic functions of the latent variables \(Y\) and \(G\), the corresponding conditional distribution is degenerate and can be defined as
\[
f_{Y^\star,H\mid Y,G}(y_i^\star,h_i\mid y,g)= \boldsymbol{1}_{I_{g}(y_{i}^\star)}(y)\boldsymbol{1}_{h_{i}}(g).
\]
 Let $
f_Y(y\mid\boldsymbol{\psi})$
denote the density (or mass) function of the latent target variable \(Y\), indexed by the parameter vector \(\boldsymbol{\psi}\in\boldsymbol{\Psi}\), and let
$f_{G\mid Y}(g\mid y;\boldsymbol{\gamma})$
denote the conditional mass function of the latent coarsening variable \(G\), where \(\boldsymbol{\gamma}\in\boldsymbol{\Gamma}\) is a vector of unknown parameters. The distribution of the coarsening levels is allowed to depend on \(Y\) itself,
reflecting the common situation in which respondents with larger latent values
are more likely to report them at a coarser resolution. Available auxiliary
information can also be included in the models for the two latent variables,
although it is omitted from the notation for simplicity.

Under this formulation, the observed-data distribution for unit \(i\) can be written as
\begin{equation}\label{eq:lik_general}
f_{Y^\star,H}(y_i^\star,h_i\mid\boldsymbol{\psi},\boldsymbol{\gamma})
=
\int_\mathcal{Y}
\sum_{g\in\mathcal G}
f_Y(y_i\mid\boldsymbol{\psi})
f_{G\mid Y}(g\mid y_i;\boldsymbol{\gamma})
f_{Y^\star,H\mid Y,G}(y_i^\star,h_i\mid y,g)
\,dy_i.    
\end{equation}
This expression represents the likelihood contribution of a single observed unit, combining the distribution of the latent variable of interest with the conditional distribution of the latent coarsening mechanism, and then marginalizing over the unobserved components.

In practice, evaluating the observed-data distribution above may be analytically cumbersome, since the latent variable \(Y\) appears both in the model for the latent response and in the model for the coarsening mechanism. A convenient approximation can be obtained by discretizing the support of \(Y\) using a step size \(\delta\).
Recall that, because of the conditional distribution
$
f_{Y^\star,H\mid Y,G}(y_i^\star,h_i\mid y_i,g_i),
$
only values of \(G\) and \(Y\) that are compatible with the observed value \(Y_i^\star=y_i^\star\) need to be considered. More precisely, for a given coarsening level \(g\in h_i\), we only consider the admissible values \(Y\in I_{g}(y_i^\star)\). Assuming that each coarsening level \(g\in\mathcal G\) is an integer multiple of \(\delta\) (excluding the notable value $g=0$), let \(K_g=g/\delta\). The idea is then to partition \(I_g(y_i^\star)\) into \(K_g\) sub-intervals, denoted by \(\{i_g^k(y_i^\star)\}_{k=1,\dots,K_g}\), such that
\[
I_g(y_i^\star)=\bigcup_{k=1}^{K_g} i_g^k(y_i^\star).
\]
These sub-intervals are defined as
\[
i_g^k(y_i^\star)
=
\left[
m_k(y_i^\star,g)-\frac{\delta}{2},\,
m_k(y_i^\star,g)+\frac{\delta}{2}
\right),
\]
where
\(
m_k(y_i^\star,g)
=
y_i^\star-\frac{g}{2}+\left(k-\frac{1}{2}\right)\delta
\)
denotes the midpoint of the \(k\)-th sub-interval. With this construction, the likelihood contribution of unit \(i\) can be approximated as
\begin{equation}\label{eq:lik_app}
f_{Y^\star,H}(y_i^\star,h_i\mid \boldsymbol{\psi},\boldsymbol{\gamma})
\approx
\sum_{g\in h_i}
\sum_{k=1}^{K_g}
f_{G\mid Y}\bigl(g\mid m_k(y_i^\star,g);\boldsymbol{\gamma}\bigr)
\int_{i_g^k(y_i^\star)}
f_Y(y_i\mid\boldsymbol{\psi})\mathrm{d} y.    
\end{equation}

To account for the sampling design, we adopt a survey-weighted pseudo-likelihood in which each sampled unit contributes to the sample likelihood with weight \(w_i\):
\[
L_s^{\,pw}(\boldsymbol{\psi},\boldsymbol{\gamma})
=
\prod_{i\in s}
\left[
f_{Y^\star,H}(y_i^\star,h_i\mid \boldsymbol{\psi},\boldsymbol{\gamma})
\right]^{\tilde{w}_i},
\]
where $\tilde{w}_i=nw_i/\sum_{i\in s}w_i$ are the re-scaled weights.
Combined with prior distributions on the model parameters, this survey-weighted pseudo-likelihood induces a pseudo-posterior distribution. In this sense, the proposed approach fits within the Bayesian pseudo-posterior framework for informative sampling, where sampling weights are incorporated through weighted likelihood contributions \citep{SavitskyToth2016}.

\subsection{The coarsening-adjusted design-based estimator}\label{sec:CA_derivation}

Starting from the pseudo-posterior distribution of the model parameters, the steps described in this section are finalized to the production of the CA design-based estimator. The first passage concerns the reconstruction of the latent sample. 
As in \citet{heitjan1990inference}, we target the posterior predictive distribution of the reconstructed latent response $\tilde{y}_i$, conditionally on the subject-specific observed value $o_i=(y^\star_i,h_i)$:
\begin{equation}
\label{eq:target_pp}
p(\tilde{y}_i \mid \mathbf{y}_s^\star,\mathbf{h}_s,o_i)
=
\int_{\boldsymbol{\Psi}}
\int_{\boldsymbol{\Gamma}}
\sum_{\tilde{g}_i\in h_i}
\pi(\tilde{y}_i,\tilde{g}_i
\mid \boldsymbol{\psi},\boldsymbol{\gamma},o_i)\,
p(\boldsymbol{\psi},\boldsymbol{\gamma}
\mid \mathbf{y}_s^\star,\mathbf{h}_s)
\,\mathrm{d}\boldsymbol{\gamma}\,
\mathrm{d}\boldsymbol{\psi}.
\end{equation}
The distribution \(
\pi(\tilde{y}_i,\tilde{g}_i
\mid \boldsymbol{\psi},\boldsymbol{\gamma},o_i)
\) denotes the joint predictive distribution of the latent response and of the
latent coarsening regime, given the observed pair $o_i$. Note that, even if the information in $o_i$ is also included in the vectors $\mathbf{y}_s^\star$ and $\mathbf{h}_s$, we include it in the conditioning part to stress its role in determining the posterior predictive of interest.

Recalling the likelihood definition \eqref{eq:lik_general}, the joint distribution, up to a proportionality constant, is defined as
\begin{equation}
\label{eq:joint_pp}
\pi(\tilde{y}_i,\tilde{g}_i
\mid \boldsymbol{\psi},\boldsymbol{\gamma}, o_i)
\propto
f_Y(\tilde{y}_i\mid\boldsymbol{\psi})\,
f_{G\mid Y}(\tilde{g}_i\mid \tilde{y}_i;\boldsymbol{\gamma})\,
\mathbf{1}\!\left\{\tilde{y}_i\in I_{\tilde{g}_i}(y_i^\star)\right\}
\mathbf{1}\!\left\{\tilde{g}_i\in h_i\right\}.
\end{equation}
By plugging this expression into the formula \eqref{eq:target_pp}, we get
\begin{equation}
\label{eq:target_pp_final}
\begin{aligned}
p(\tilde{y}_i \mid \mathbf{y}_s^\star,\mathbf{h}_s,o_i)
\propto
\int_{\boldsymbol{\Psi}}
\int_{\boldsymbol{\Gamma}}&
f_Y(\tilde{y}_i\mid\boldsymbol{\psi})
\left(\sum_{\tilde{g}_i\in h_i}
f_{G\mid Y}(\tilde{g}_i\mid \tilde{y}_i;\boldsymbol{\gamma})
\mathbf{1}\!\left\{
\tilde{y}_i\in I_{\tilde{g}_i}(y_i^\star)
\right\}\right)\times\\
&\quad p(\boldsymbol{\psi},\boldsymbol{\gamma}
\mid \mathbf{y}_s^\star,\mathbf{h}_s)
\,\mathrm{d}\boldsymbol{\gamma}\,
\mathrm{d}\boldsymbol{\psi}.
\end{aligned}
\end{equation}

Defining $\tilde{\mathbf{y}}_s=\{\tilde{Y}_i\}_{i\in s}$ as the collection of latent variables that constitutes the reconstructed sample and
$\mathbf{o}_{s}=\left\{(y^\star_i,h_i)\right\}_{i\in s}$ as the set
of observed pairs, we are finally interested in the posterior distribution
of the design-based estimator
$T(\tilde{\mathbf{y}}_s,\mathbf{w}_s)|\mathbf{y}_s^\star,\mathbf{h}_s,\mathbf{o}_s$.
It is important to point out that we are dealing with two sources of variation: the uncertainty due to the use of an
imputation model and the uncertainty due to the sampling design \citep{isaki1982survey}. We denote
expected values and variances related to the first source by
$\mathbb{E}_m$ and $\mathbb{V}_m$, while $\mathbb{E}_p$ and
$\mathbb{V}_p$ denote the corresponding operators with respect to the
sampling design.

As a CA estimator, the posterior mean of this functional
appears to be a reasonable choice. More explicitly, one may write
\begin{equation}\label{eq:stimatore}
\hat{\theta}_\text{CA}
=
\mathbb{E}_m\left[
\mathbb{E}_p\left[
T(\tilde{\mathbf{y}}_s,\mathbf{w}_s)
\mid \tilde{\mathbf{y}_s}
\right]
\mid \mathbf{y}_s^\star,\mathbf{h}_s,\mathbf{o}_s
\right]=
\mathbb{E}_m\left[
T(\tilde{\mathbf{y}}_s,\mathbf{w}_s)
\mid \mathbf{y}_s^\star,\mathbf{h}_s,\mathbf{o}_s
\right],
\end{equation}
as once we condition on the latent data
\(\tilde{\mathbf{y}}_s\), the observed sample \(s\) and the associated
weights \(\mathbf{w}_s\) are fixed and $\mathbb{E}_p\left[
T(\tilde{\mathbf{y}}_s,\mathbf{w}_s)
\mid \tilde{\mathbf{y}}_s
\right]
=
T(\tilde{\mathbf{y}}_s,\mathbf{w}_s).$
The distinction between the two sources of variation becomes relevant
when evaluating the uncertainty of the estimator. Conditional on
\(\tilde{\mathbf{y}}_s\), the latent variables are treated as fixed and the
remaining variability is due to the sampling design. Therefore, by the law
of total variance,
\begin{equation}\label{eq:var}
\begin{aligned}
\mathbb{V}[\hat{\theta}_\text{CA}]
&=
\mathbb{E}_m\left[
\mathbb{V}_p\left[
T(\tilde{\mathbf{y}}_s,\mathbf{w}_s)
\mid \tilde{\mathbf{y}}_s
\right]
\mid \mathbf{y}_s^\star,\mathbf{h}_s,\mathbf{o}_s
\right] +
\mathbb{V}_m\left[
T(\tilde{\mathbf{y}}_s,\mathbf{w}_s)
\mid \mathbf{y}_s^\star,\mathbf{h}_s,\mathbf{o}_s
\right].%\\
%&=\mathbb{E}_m\left[
%\mathbb{V}_p\left[
%T(\tilde{\mathbf{y}}_s,\mathbf{w}_s)
%\right]
%\mid \mathbf{y}_s^\star,\mathbf{h}_s,\mathbf{o}
%\right] +
%\mathbb{V}_m\left[
%T(\tilde{\mathbf{y}}_s,\mathbf{w}_s)
%\mid \mathbf{y}_s^\star,\mathbf{h}_s,\mathbf{o}
%\right].
\end{aligned}
\end{equation}
The first term accounts for the uncertainty induced by the sampling design,
averaged over the posterior predictive distribution of the latent data.
The second term accounts for the posterior predictive uncertainty induced
by the imputation model, through the variability of the estimator across
different reconstructed versions of the sample. The treatment of uncertainty in this decomposition shares the same rationale as
multiple imputation methods, where variability is separated into within- and
between-imputation components \citep{rubin1996multiple,murray2018multiple}.
Therefore, the percentage
contribution of the second term to the total variance,
\begin{equation}\label{eq:frac}
\eta_\text{C}
=
\frac{
\mathbb{V}_m\left[
T(\tilde{\mathbf{y}}_s,\mathbf{w}_s)
\mid \mathbf{y}_s^\star,\mathbf{h}_s,\mathbf{o}_s
\right]
}{
\mathbb{V}[\hat{\theta}_\text{CA}]
}\times 100,
\end{equation}
can be interpreted as an indicator of the fraction of uncertainty
attributable to coarsening. Values close to zero suggest that the
uncertainty is mainly driven by the sampling design, while larger values
indicate a stronger impact of the coarsening mechanism and of the
associated imputation model.

By replacing the design-based variance with a suitable estimator, the variance
of the CA estimator can be estimated as
\begin{equation}\label{eq:est_var}
\widehat{V}[\hat{\theta}_{\mathrm{CA}}]
=
\mathbb{E}_m\left[
\widehat{V}_p\left[
T(\tilde{\mathbf{y}}_s,\mathbf{w}_s)\mid \tilde{\mathbf{y}}_s
\right]
\mid \mathbf{y}_s^\star,\mathbf{h}_s,\mathbf{o}_s
\right]
+
\mathbb{V}_m\left[
T(\tilde{\mathbf{y}}_s,\mathbf{w}_s)
\mid \mathbf{y}_s^\star,\mathbf{h}_s,\mathbf{o}_s
\right].
\end{equation}
Plugging this quantity into \eqref{eq:frac} gives the estimated fraction
\(\widehat{\eta}_C\). The same variance estimator is also used to construct the
approximate \((1-\alpha)\)-level confidence interval
\(
\hat{\theta}_{\mathrm{CA}}
\pm
z_{1-\alpha/2}
\sqrt{\widehat{V}[\hat{\theta}_{\mathrm{CA}}]} .
\)

\subsection{Computational aspects}

The steps described in Sections~\ref{sec:model} and~\ref{sec:CA_derivation} involve some computational aspects that deserve further
discussion.
First, the model presented in Section~\ref{sec:model} can be fitted
once suitable model distributions have been specified for both the latent
variable \(Y\) and the coarsening behaviour. In particular, exploiting the
discretized formulation in Equation~\eqref{eq:lik_app}, the model can be
implemented using standard probabilistic programming languages such as
\texttt{Stan} \citep{carpenter2017stan}. Within this framework, Hamiltonian Monte Carlo (HMC), a specific Markov chain
Monte Carlo (MCMC) method, can be used to obtain \(B\) draws from the pseudo-posterior
distributions of the model parameters, namely
\(p(\boldsymbol{\psi}\mid \mathbf{z}_s^\star,\mathbf{h}_s)\) and
\(p(\boldsymbol{\gamma}\mid \mathbf{z}_s^\star,\mathbf{h}_s)\).

The subsequent step consists in generating values from the posterior predictive
distribution of the latent variable for each sampled unit \(i\), conditionally
on \(o_i\).
Given the posterior draws
\((\boldsymbol{\psi}^{(b)},\boldsymbol{\gamma}^{(b)})\),
\(b=1,\ldots,B\), predictive replicates are generated from \eqref{eq:target_pp_final} using a grid sampler, exploiting the fact that, conditional on the observed coarsened value $y_i^\star$ and on the compatible coarsening regimes $h_i$, the support of the latent value $\tilde y_i$ is restricted to the finite union of compatibility intervals $\bigcup_{g \in h_i} I_g(y_i^\star)$. More specifically, a constant step size is chosen and, for
each unit \(i\), a deterministic grid
\(
\left\{y_{i1},\ldots,y_{iM_i}\right\}
\)
is defined over the union of the intervals \(I_g(y_i^\star)\), with
\(g\in h_i\). For each posterior draw \(b\) and each grid point \(y_{im}\), the
unnormalized weight is
\begin{equation}
\label{eq:grid_weights}
v_{im}^{(b)}
=
f_Y(y_{im}\mid\boldsymbol{\psi}^{(b)})
\sum_{g\in h_i}
f_{G\mid Y}(g\mid y_{im};\boldsymbol{\gamma}^{(b)})
\mathbf{1}\!\left\{
y_{im}\in I_g(y_i^\star)
\right\}.
\end{equation}
After normalizing the weights as
\(
\bar{v}_{im}^{(b)}
=
v_{im}^{(b)}
/
\sum_{\ell=1}^{M_i} v_{i\ell}^{(b)}
\),
the predictive replicates \(\tilde{y}_i^{(b)}\) are generated, obtaining \(B\)
generated samples from the latent variable:
\[
\tilde{\mathbf y}_s^{(b)}
=
\left\{
\tilde y_i^{(b)}
\sim
p(\tilde y_i
\mid
\mathbf y_s^\star,\mathbf h_s,o_i)
\right\}_{i\in s},
\qquad b=1,\ldots,B.
\]

For each Monte Carlo (MC) sample of the latent variable
\(\tilde{\mathbf y}_s^{(b)}\), we compute the design-based estimate
\(\hat\theta^{(b)}=T(\tilde{\mathbf y}_s^{(b)},\mathbf w_s)\) and the
corresponding estimate of the design-based variance
\(
\widehat V_p(\hat{\theta})^{(b)}
=
\widehat V_p\left[
T(\tilde{\mathbf y}_s^{(b)},\mathbf w_s)
\right]
\).
Then, the MC mean provides the numerical estimate of
\eqref{eq:stimatore}:
\[
\hat\theta_{\mathrm{CA}}
=
\mathbb E\left[
T(\tilde{\mathbf y}_s,\mathbf w_s)
\mid \mathbf y_s^\star,\mathbf h_s
\right]
\approx
\frac{1}{B}\sum_{b=1}^B \hat\theta^{(b)} ,
\]
while the numerical evaluation of the total variance estimate \eqref{eq:est_var} is given by
\[
\widehat V(\hat\theta_{\mathrm{CA}})
\approx
\frac{1}{B}\sum_{b=1}^B \widehat V_p(\hat{\theta})^{(b)}
+
\frac{1}{B-1}\sum_{b=1}^B
\left(\hat\theta^{(b)}-\hat\theta_{\mathrm{CA}}\right)^2.
\]

\section{An Application on PASSI data: smoking behaviour}\label{sec:application}

The general procedure described in Section \ref{sec:procedure} must be translated into a concrete modeling strategy tailored to the specific features of the data. In this application, the target variable is the average number of daily smoked cigarettes. Note that, by definition, this response has a semi-continuous structure, since the value \(0\) identifies an individual who is not a daily smoker. For this reason, the observed response can be expressed as the product of two distinct variables:
\(
Y^\star = U \cdot Z^\star,
\)
where \(U=1\) denotes a daily smoker and \(U=0\) a non-smoker. Consequently, \(Z^\star>0\) represents the average number of cigarettes smoked per day by daily smokers and is relevant only when \(U=1\).

In this application, we assume that \(U\) is correctly recorded, whereas \(Z^\star\) is a discrete and coarsened version of the latent continuous variable \(Z\in\mathbb{R}^+\). Moreover, the covariate vector \(\mathbf{x}_i\) includes a categorical variable for stratum, defined by the combination of gender and 3 age groups (18-34, 35-49, 50-69), which are incorporated as predictors in the model. The first modeling choice concerns the set of coarsening levels, for which we assume
\(
\mathcal{G}=\{1,5,10\}.
\)
These values are interpreted as ordered categories of the coarsening mechanism. The probability associated with each coarsening level is defined as \(\mathbb{P}(G=g\mid z,\boldsymbol{\gamma})=\lambda_g(z,\boldsymbol{\gamma})\), and a proportional-odds model \citep{wang2008modeling} is assumed to relate these probabilities to the latent variable \(Z\), thereby specifying the structure of \(f_{G\mid Z}(g\mid z;\boldsymbol{\gamma})\):
\begin{equation}\label{eq:probsG}
\begin{cases}
    \lambda_1(z,\boldsymbol{\gamma}) = \text{expit}(\gamma_{01}+\gamma_1\log z),\\
    \lambda_5(z,\boldsymbol{\gamma}) = \text{expit}(\gamma_{02}+\gamma_1\log z)-\text{expit}(\gamma_{01}+\gamma_1\log z), \\
    \lambda_{10}(z,\boldsymbol{\gamma}) = 1-\text{expit}(\gamma_{02}+\gamma_1\log z)
\end{cases}
\end{equation}
with $\gamma_{02} >\gamma_{01}$.

To complete the modeling framework, a distributional assumption for the latent variable \(Z\) is required. Following the considerations in \citet{gardini2026coarsened}, we consider two alternative models. The simpler specification assumes a Lognormal (LN) distribution for \(Z\):
\begin{equation}
    Z_i \mid \boldsymbol{\psi} \sim \mathcal{LN}\left(\beta_0^\mu+\mathbf{x}_i^\top\boldsymbol{\beta}^\mu,\sigma^2\right).
\end{equation}
To better capture the distributional shape of the average number of cigarettes smoked, a two-component mixture model (LNM) may be preferable:
\begin{equation}\label{eq:mixt}
    Z_i \mid l_i,\boldsymbol{\psi} \sim \boldsymbol{1}_{\{1\}}(l_i)\mathcal{LN}\left(\beta_{01}^\mu+\mathbf{x}_i^\top\boldsymbol{\beta}^\mu,\sigma_1^2\right)+\boldsymbol{1}_{\{2\}}(l_i)\mathcal{LN}\left(\beta_{02}^\mu+\mathbf{x}_i^\top\boldsymbol{\beta}^\mu,\sigma_2^2\right),
\end{equation}
where \(l_i\in\{1,2\}\) is a latent allocation variable. For identifiability, we impose \(\beta_{01}<\beta_{02}\), and model the probability of the mixture label as
$
\mathbb{P}(l_i=1\mid\mathbf{x}_i)=\text{expit}\left(\beta_0^\pi+\mathbf{x}_i^\top\boldsymbol{\beta}^\pi\right).
$
The final approximated likelihood \eqref{eq:lik_app} is obtained by setting the sub-interval width as $\delta=1$.

These model components enter the survey-weighted pseudo-likelihood and consequently define the corresponding pseudo-posterior distribution once prior distributions are assigned to the model parameters. In particular, we opt for a standard weakly-informative prior setting, assuming independent $\mathcal{N}(0,2.5^2)$ priors of coefficients for standardised covariates and $\mathcal{N}^+(0,2.5^2)$ priors for scale parameters.

\subsection{Diagnostics for the fitted models}

\begin{figure}
    \centering
    \includegraphics[width=1\linewidth]{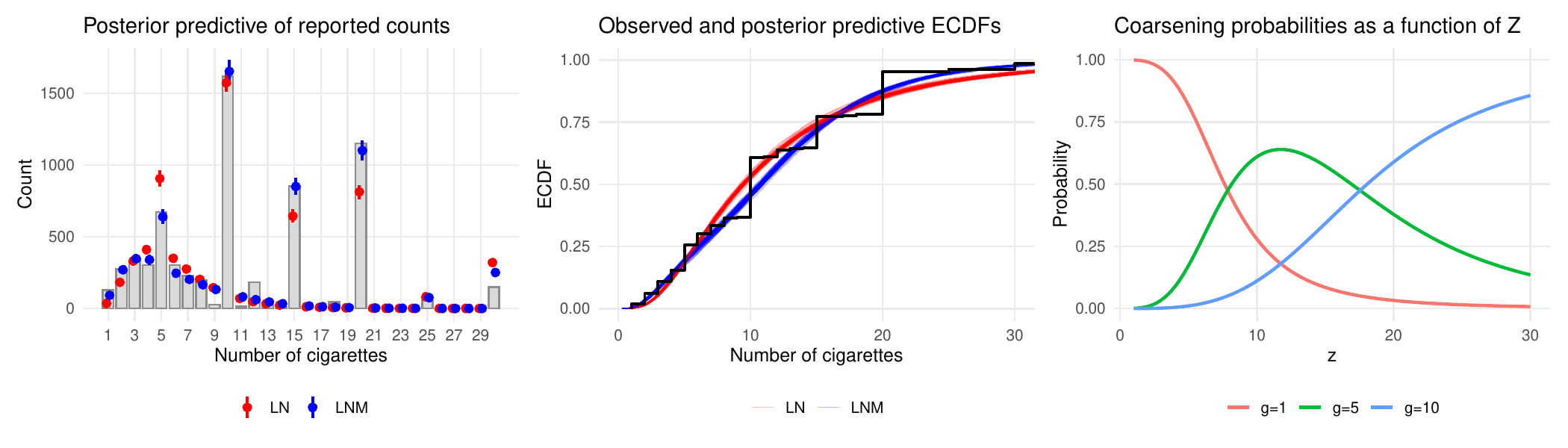}
\caption{Model diagnostics for the LN and LNM models. The left and central panels show posterior predictive checks based on reported counts and ECDFs, respectively, while the right panel reports the fitted coarsening probabilities under the LNM model as a function of $Z$.}
    \label{fig:fig_models_appl}
\end{figure}

Before moving to the results obtained for the estimates of the target population quantities, it is useful to comment on the fitted models. A first point concerns the comparison between the LN and LNM models, which is assessed against the observed data through two posterior predictive checks, reported in Figure~\ref{fig:fig_models_appl}.
The bar plot in the left panel compares the coarsened samples generated from the posterior predictive distributions of the models with the observed distribution of \(Z^\star\). The points represent the average counts in the replicated datasets, while the vertical bars denote 95\% credible intervals. This plot suggests that the LNM model improves the performance of the LN model, as it can reproduce the observed coarsened data more closely.
The central panel compares the empirical cumulative distribution functions (ECDFs) of the continuous latent outcome $Z$ under the two models with the ECDF of $Z^\star$. Also in this case, the ECDFs obtained under the LNM model appear to provide a better interpolation of the ECDF of the observed data.
Finally, the right-hand panel shows the fitted probabilities of the coarsening regimes under the LNM model, focusing on how they vary with $Z$. As expected, the probability of rounding to the nearest unit decreases rapidly as $Z$ increases. Conversely, for larger values of $Z$, individuals tend to report values heaped at multiples of 10.

\begin{figure}
    \centering
    \includegraphics[width=.9\linewidth]{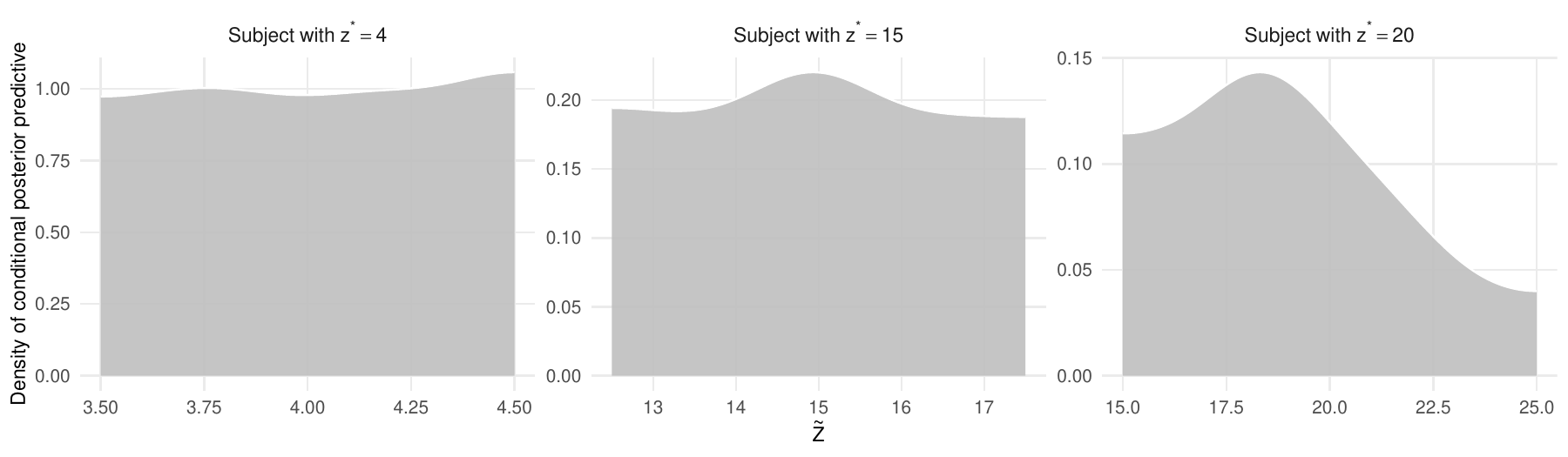}
    \caption{Posterior predictive distributions of the latent variable $\tilde{Z}$ under the LNM model, conditional on selected observed coarsened values $z^\star$.}
    \label{fig:pred_latent_appl}
\end{figure}

Focusing on the LNM model, Figure~\ref{fig:pred_latent_appl} reports some posterior predictive distributions conditional on selected observed coarsened values, as described in Section~\ref{sec:CA_derivation}. The three kernel densities illustrate how the uncertainty on the latent value increases as the set of compatible coarsening regimes becomes larger. In the left panel, where $h_i=\{1\}$, only the unit-rounding regime is compatible with the observed value. In the central panel, where $h_i=\{1,5\}$, the unit-rounding component is combined with the distribution over the interval of width 5 around the observed value. Finally, in the right panel, where $h_i=\{1,5,10\}$, the interval of width 10 is also included, leading to a considerable increase in posterior uncertainty.

\subsection{Heavy smokers proportion}
In the first analysis, we focus on an indicator widely used to monitor smoking intensity in a population, namely the proportion of so-called heavy smokers, defined as individuals who smoke, on average, 20 or more cigarettes per day \citep{eurostat_tobacco_2022}.

To this aim, we consider the whole target population $\mathcal{U}$, together with its partition into administrative regions (NUTS-2), denoted by
$\{\mathcal{U}_1,\dots,\mathcal{U}_D\}$, with $\bigcup_d \mathcal{U}_d=\mathcal{U}$ and $|\mathcal{U}_d|=N_d$. The target quantities are defined at the population level as
\begin{equation}\label{eq:prop_def}
\pi^{\text{HS}}=
\frac{\sum_{i\in\mathcal{U}}\boldsymbol{1}_{[20,+\infty)}(y_i)}{N}
\quad \text{and}\quad
\pi^{\text{HS}}_d=
\frac{\sum_{i\in\mathcal{U}_d}\boldsymbol{1}_{[20,+\infty)}(y_i)}{N_d},
\end{equation}
for the whole population and for the $d$-th sub-population, respectively.

A natural naive direct estimator, based only on the observed coarsened values $\mathbf{y}_s^\star$, is the Hájek-type estimator
\[
\widehat{\pi}^{\text{HS}}_{\text{Naive}}=
\frac{\sum_{i\in s}w_i\boldsymbol{1}_{[20,+\infty)}(y_i^\star)}
{\sum_{i\in s}w_i}
\quad \text{and}\quad
\widehat{\pi}^{\text{HS}}_{d,\text{Naive}}=
\frac{\sum_{i\in s_d}w_i\boldsymbol{1}_{[20,+\infty)}(y_i^\star)}
{\sum_{i\in s_d}w_i},
\]
which can be implemented using the \texttt{R} package \texttt{survey} \citep{lumley2004survey}. The package provides variance estimates and confidence limits
constructed on the logit scale (see Algorithm~S1 in the Supplementary Material),
ensuring that the resulting interval lies within \((0,1)\).

To account for the fact that only a coarsened version of the latent target
variable is observed, we use the CA estimators discussed
in Section~\ref{sec:procedure}, exploiting predictions from the models described
above. Both the LN and LNM specifications are considered, leading to the estimates \(\widehat{\pi}^{\text{HS}}_{\text{CA-LN}}\) and
\(\widehat{\pi}^{\text{HS}}_{\text{CA-LNM}}\), together with their corresponding
variance estimates. To obtain confidence intervals constrained to lie in
\((0,1)\), we adopt Algorithm~S2 in the Supplementary Material, which exploits the
logit transformation.

To highlight the potential issues of the naive estimator, which ignores the coarsening mechanism, we also consider the quantities
\begin{equation}
\pi^{\text{HS}^{\prime}}=
\frac{\sum_{i\in\mathcal{U}}\boldsymbol{1}_{(20,+\infty)}(y_i)}{N}
\quad \text{and}\quad
\pi^{\text{HS}^{\prime}}_d=
\frac{\sum_{i\in\mathcal{U}_d}\boldsymbol{1}_{(20,+\infty)}(y_i)}{N_d}.
\end{equation}
They differ from $\pi^{\text{HS}}$ and $\pi^{\text{HS}}_d$ only in the exclusion of the threshold value 20. Since $Y$ is theoretically continuous, the difference between the two population quantities should be negligible. In contrast, when using the observed coarsened values $Y^\star$, the mass accumulated at the heaping value 20 may induce non-negligible differences between the corresponding naive estimators.

\begin{figure}
    \centering
\includegraphics[width=.9\linewidth]{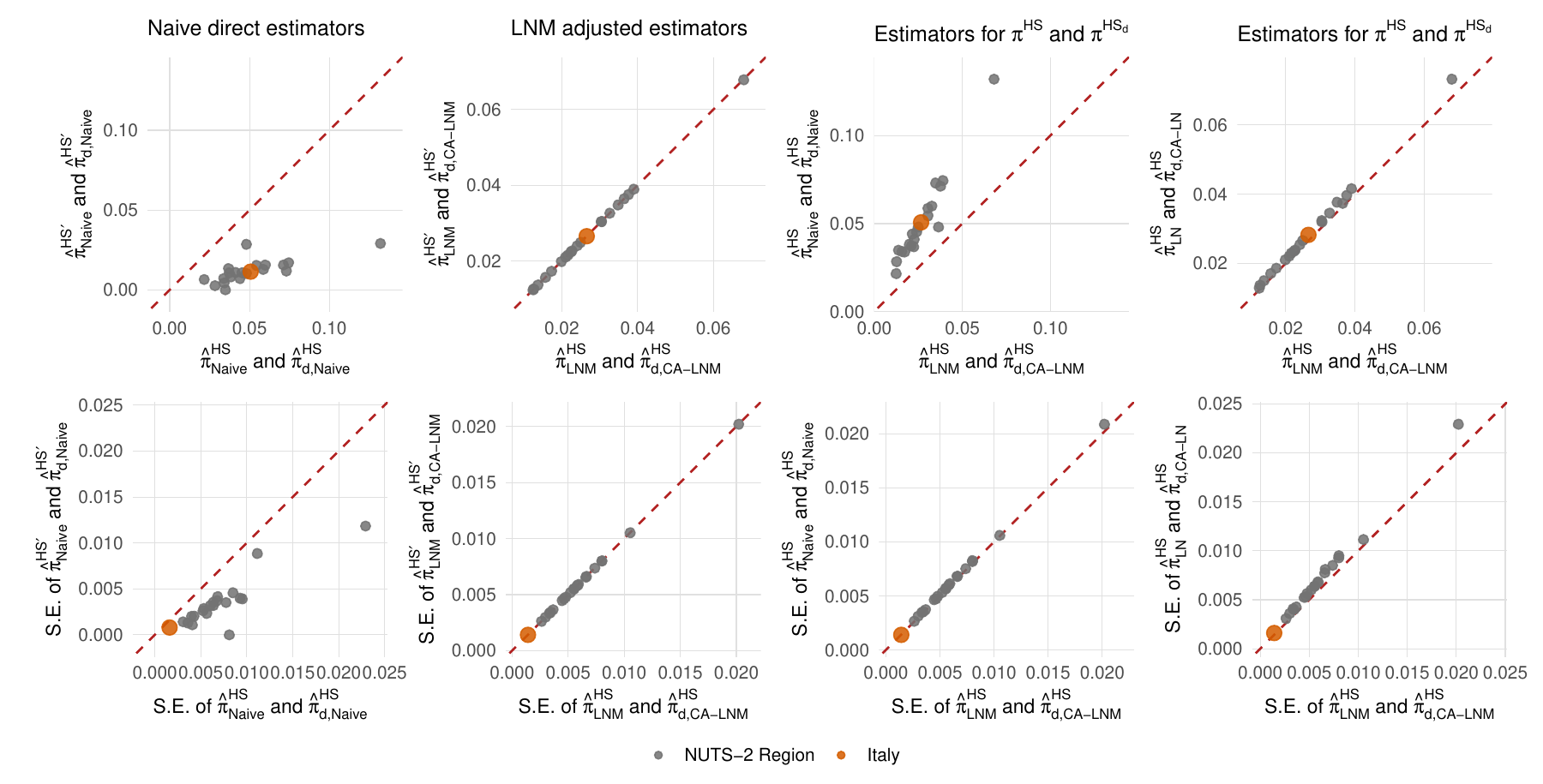}
    \caption{Point estimates and standard errors (S.E.) under the naive direct and CA strategies. The panels compare estimates for $\pi^{\text{HS}}$ and $\pi^{\text{HS}^{\prime}}$, and assess the sensitivity of the CA estimates to the LN and LNM models.}
    \label{fig:cfr_1}
\end{figure}
Figure~\ref{fig:cfr_1} compares point estimates and standard errors across the
different estimation strategies. A first result concerns the comparison between
the estimates of the two quantities \(\pi^{\text{HS}}\) and
\(\pi^{\text{HS}^{\prime}}\). For the naive estimator, the differences
are remarkable: the estimates of \(\pi^{\text{HS}^{\prime}}\) are substantially
lower than those of \(\pi^{\text{HS}}\), as expected from the data distribution
shown in Figure~\ref{fig:esplorativa}. Indeed, the value 20 represents a
relevant heaping point, and including it in the definition of the indicator
artificially inflates the estimated HS proportion. Conversely, the
CA-LNM estimates do not show relevant
differences between \(\pi^{\text{HS}}\) and \(\pi^{\text{HS}^{\prime}}\).
Focusing on the policy-relevant indicator \(\pi^{\text{HS}}\), we observe a
clear difference between the naive strategy and the proposed
CA strategy, with higher values obtained under the naive
estimator. The differences in the point estimates are not accompanied by substantial changes in the estimated standard errors, although the latter are not directly comparable because of the differences in the magnitude of the estimates. A final relevant result
concerns the robustness of the CA strategy with respect to the
model choice. Indeed, despite the marked differences between the LN and LNM
models highlighted in the model comparison stage, the two approaches lead to
very similar point estimates. 

\begin{figure}
    \centering
    \includegraphics[width=1\linewidth]{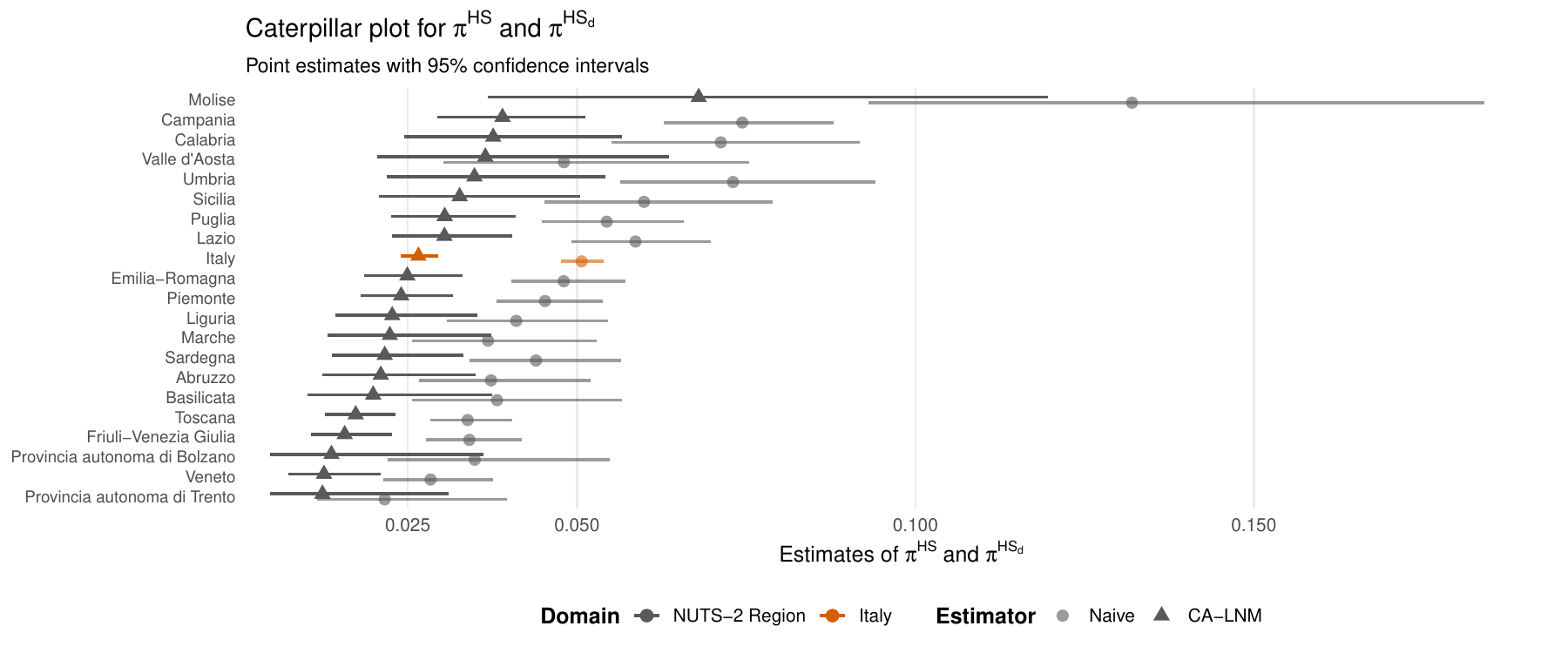}
    \caption{Caterpillar plot of naive direct and CA-LNM estimates of the HS prevalence for Italy and NUTS-2 regions. Points denote estimated proportions and horizontal bars represent 95\% confidence intervals.}
    \label{fig:caterpillar}
\end{figure}

Figure~\ref{fig:caterpillar} complements the comparison in
Figure~\ref{fig:cfr_1} by reporting the estimates separately for each NUTS-2
region. While the previous figure highlighted the overall distribution of the
regional estimates across methods, the caterpillar plot makes it possible to
link each estimate to its corresponding territory and to compare the naive and
CA-LNM strategies within each region. The figure also displays the
associated confidence intervals, allowing the magnitude of the correction to be
evaluated relative to the uncertainty of the estimates.
 For Italy and for some NUTS-2 regions, the
confidence intervals obtained under the naive and CA-LNM strategies do
not overlap. Therefore, the correction is not only visible in terms of
point estimates, but also relevant from an inferential perspective. Focusing on the CA intervals, they substantially overlap for many regions, suggesting
that small differences between regional point estimates should be interpreted
with caution, since they may be compatible with sampling and coarsening
uncertainty.

Overall, these results suggest that ignoring coarsening may overstate the
prevalence of HS and, in some cases, the apparent magnitude of
regional differences. This may affect policy priorities, territorial
benchmarking, and the allocation of prevention efforts, which could partly
reflect artefacts of the reporting process rather than genuine differences in
smoking behaviour. The CA-LNM estimates therefore provide a more
appropriate basis for monitoring and targeting prevention policies.

%\textcolor{red}{Gaia: Da mettere commento su caterpillar plot}

\subsection{Investigating the population of daily smokers}\label{sec:smokers}

In this section, the focus is restricted to $\mathcal{U}^+$, namely the population of daily smokers. In addition to the overall population, we consider a set of sub-populations $\mathcal{U}^+_1,\dots,\mathcal{U}^+_D$, obtained by crossing macro-region (North, Centre, South with Islands), gender, and age class (18--34, 35--49, and 50--69). This yields $D=18$ sub-populations.

\begin{figure}
    \centering
    \includegraphics[width=.9\linewidth]{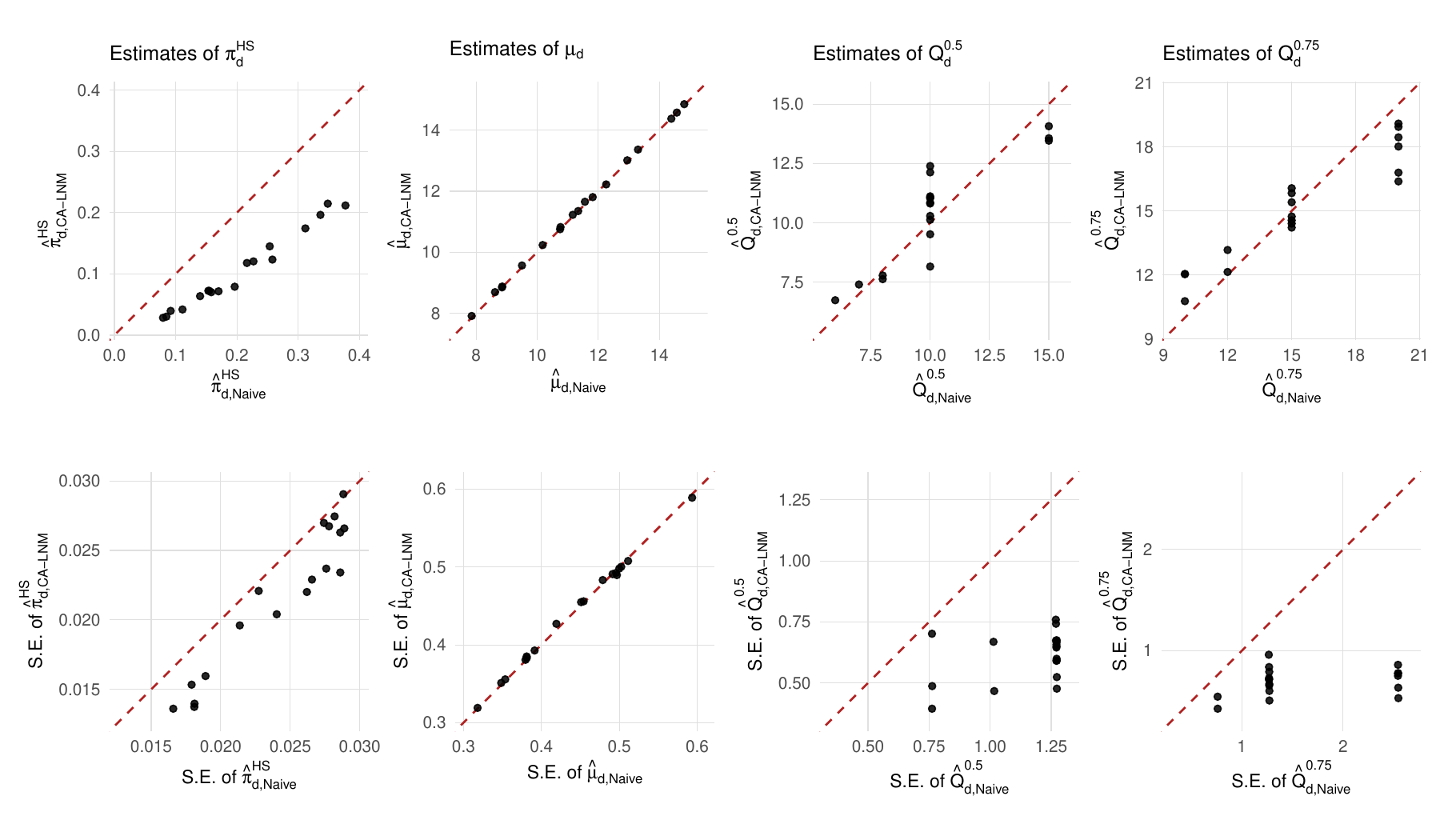}
    \caption{Comparison between naive and CA-LNM estimates for daily smokers across macro-region, gender and age-class domains. The upper panels report point estimates for the HS proportion, mean, median and third quartile; the lower panels report the corresponding standard errors.}
    \label{fig:placeholder_estcomp}
\end{figure}

Besides the HS proportion, $\pi^{\mathrm{HS}}$, defined as in \eqref{eq:prop_def}, with $\mathcal{U}$ replaced by $\mathcal{U}^+$, this section also targets the mean and selected quantiles of the average number of cigarettes smoked per day. The population mean is defined as
$
\mu =
\frac{1}{N^+}\sum_{i\in\mathcal{U}^+} y_i,
$
and, for subpopulation $d$,
$
\mu_d =
\frac{1}{N_d^+}\sum_{i\in\mathcal{U}_d^+} y_i.
$
The corresponding naive Hájek-type estimators are
\[
\widehat{\mu}_{\mathrm{Naive}} =
\frac{\sum_{i\in s^+} w_i y_i^\star}
{\sum_{i\in s^+} w_i},
\qquad
\widehat{\mu}_{d,\mathrm{Naive}} =
\frac{\sum_{i\in s_d^+} w_i y_i^\star}
{\sum_{i\in s_d^+} w_i}.
\]

Quantiles represent an additional quantity of interest. Let the finite-population cumulative distribution function be defined as
\(
F(t)=
\frac{1}{N^+}
\sum_{i \in \mathcal{U}^+}
\boldsymbol{1}_{(-\infty,t]}(y_i).
\)
The corresponding $p$-th quantile is defined as the generalized inverse of the cumulative distribution function,
\(
Q^p = F^{-1}(p)
= \inf_t \{F(t)\geq p\}.
\)
Under a complex survey design, quantiles can be estimated using the \texttt{svyquantile} function from the \texttt{survey} package. In its default setting, this estimator is defined as the mathematical inverse of the estimated cumulative distribution function. The naive weighted estimator of the CDF is therefore
\[
\widehat{F}_{\mathrm{Naive}}(t)=
\frac{
\sum_{i\in s^+} w_i
\boldsymbol{1}_{(-\infty,t]}(y_i^\star)
}{
\sum_{i\in s^+} w_i
},
\]
which leads to the naive quantile estimator
\[
\widehat{Q}^p_{\mathrm{Naive}}
=
\inf_t
\{\widehat{F}_{\mathrm{Naive}}(t)\geq p\}.
\]
Analogous definitions apply to each subpopulation $\mathcal{U}_d^+$, yielding $F_d(t)$, $Q_d^p$, $\widehat{F}_{d,\mathrm{Naive}}(t)$, and $\widehat{Q}^p_{d,\mathrm{Naive}}$.  CA counterparts are
obtained by applying the procedure proposed in Section~\ref{sec:procedure}; in
the following, we focus on the results obtained under the LNM model.

Figure~\ref{fig:placeholder_estcomp} and Table~\ref{tab:italy_daily_smokers}
compare the naive and CA-LNM estimators when the analysis is restricted to
daily smokers. For the HS proportion, all domain-level CA-LNM
estimates are lower than the corresponding naive estimates, consistently with
the results obtained for the whole population. At the national level, the
estimated proportion decreases from 0.23 to 0.12, with non-overlapping
confidence intervals. By contrast, point estimates and standard errors for the mean are very similar under the two
strategies. This suggests
that, in this application, the mean is relatively robust to coarsening,
plausibly because positive and negative rounding errors tend to compensate when
averaged. Thus, coarsening does not affect all indicators in the same way:
threshold-based indicators are highly sensitive to heaping at policy-relevant
cut-points, whereas averages may be much less affected.

\begin{table}
\centering
\caption{Naive and CA-LNM estimates for daily smokers in Italy. Point estimates, standard errors and 95\% confidence intervals are reported.}
\label{tab:italy_daily_smokers}
\begin{tabular}{lcccccc}
\toprule
& \multicolumn{3}{c}{Naive} & \multicolumn{3}{c}{CA-LNM} \\
\cmidrule(lr){2-4} \cmidrule(lr){5-7}
Quantity 
& Estimate & S.E. & 95\% CI 
& Estimate & S.E. & 95\% CI \\
\midrule
$\pi^{\mathrm{HS}}$ 
& 0.23 & 0.01 & [0.21, 0.24] 
& 0.12 & 0.01 & [0.11, 0.13] \\

$\mu$ 
& 11.80 & 0.12 & [11.57, 12.03] 
& 11.83 & 0.12 & [11.60, 12.06] \\

$Q^{0.50}$ 
& 10.00 & 0.51 & [10.00, 12.00] 
& 10.96 & 0.20 & [10.62, 11.30] \\

$Q^{0.75}$ 
& 15.00 & 1.28 & [15.00, 20.00] 
& 15.99 & 0.18 & [15.66, 16.31] \\
\bottomrule
\end{tabular}
\end{table}

The impact of coarsening becomes evident again for the quantiles. The naive
estimates of the median and upper quartile are strongly influenced by the
discreteness of the reported distribution and tend to be locked at preferred
values, such as 10, 15, and 20 cigarettes per day. The CA-LNM estimates instead
reconstruct a continuous latent distribution, producing
differentiated estimates across domains. At the national level, the median
increases from 10.00 to 10.96 cigarettes per day, while the third quartile
increases from 15.00 to 15.99. The corresponding confidence intervals are also
narrower under the CA-LNM strategy, especially for the third quartile. This suggests that the naive uncertainty partly
reflects the artificial jumps of the empirical distribution at heaping points,
rather than sampling variability alone.
Confidence intervals for quantiles are computed using a Woodruff-type
construction \citep{woodruff1952confidence,sitter2001note}, based on confidence
limits for the estimated distribution function and subsequent inversion of these
limits (see Algorithm~S3). In the CA-LNM case, the same logic is applied to each
posterior predictive replicate of the latent sample, and the final endpoints are
obtained by averaging the replicate-specific limits (see Algorithm~S4). This
keeps the interval construction coherent with the survey-weighted quantile
estimator while incorporating the uncertainty induced by the reconstruction of
the latent values.

\begin{figure}
    \centering
    \includegraphics[width=0.65\linewidth]{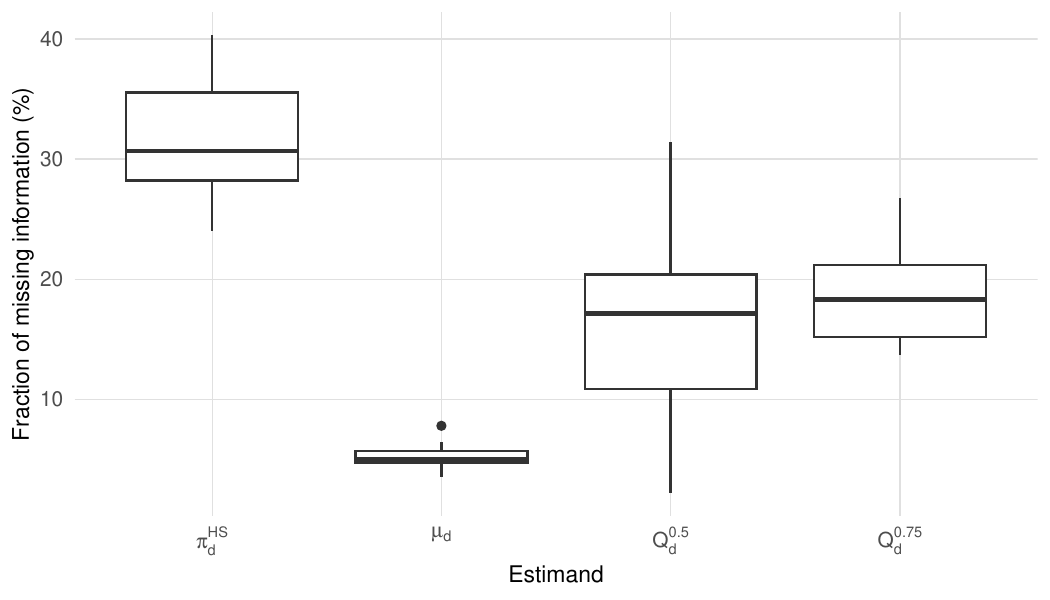}
    \caption{Fraction of missing information attributable to coarsening for the HS proportion, mean, median and third quartile across domains of daily smokers.}
    \label{fig:placeholder_boxplot:fracrion}
\end{figure}

Finally, Figure~\ref{fig:placeholder_boxplot:fracrion} reports the estimated
fractions of uncertainty attributable to coarsening, $\widehat{\eta}_C$. The differences across indicators mirror the discrepancies
observed in the point estimates. The mean shows a limited loss of information
due to coarsening, around 5\%, whereas the HS proportion displays
a much larger fraction, around 30\%, followed by the quantile-based quantities.
These results reinforce the need to assess the effect of coarsening separately
for each target indicator, rather than assuming a uniform impact across
estimands.
The additional disaggregated results reported in Section S2 of the Supplementary Material show that the CA-LNM estimates preserve clear demographic patterns. Across
macro-regions, smoking intensity and the prevalence of HS are
generally higher among men than among women and tend to increase with age. This
suggests that the adjustment for coarsening does not remove substantive
heterogeneity; rather, it reduces the component of the observed differences
that is attributable to the reporting process.

\section{Simulation study}\label{sec:sim}

To investigate the nature of the notable discrepancies between the naive design-based estimators and the proposed CA strategy observed in the application of Section \ref{sec:application}, we carried out a simulation study based on a synthetic finite population from which repeated samples are subsequently drawn. The population-generating mechanism is designed to mimic the setting observed in the sub-population of daily smokers analysed in Section \ref{sec:smokers}. Specifically, the finite population is generated according to a Lognormal mixture model calibrated on the empirical results obtained in the application. The overall population size is set to $N=17{,}500$, subdivided into $D=20$ sub-populations with sizes $N_d\in\{500,750,1000,1250\}$, with five sub-populations for each size.

As a covariate, which also determines the population strata, we first generate a categorical random variable with three equally likely classes. This variable is then represented through dummy coding, yielding a two-dimensional covariate vector $\mathbf{x}_{id}$ for each unit $i$ in sub-population $d$. Differences across sub-populations are introduced through random terms independently generated as
$u_d\sim\mathcal{N}(0,0.1^2)$.
Since the population-generating process is based on the mixture model in \eqref{eq:mixt}, a latent component label $l_{id}\in\{1,2\}$ is first sampled with probability
\[
\mathbb{P}(l_{id}=1)=\operatorname{expit}(-0.07-0.15x_{id,1}-0.25x_{id,2}).
\]
Conditional on the sampled component label, the unit-level target response $y_{id}$ is generated from
\[
y_{id}\mid l_{id} \sim 
\mathcal{LN}\left(
b_{0l_{id}}+0.08x_{id,1}+0.17x_{id,2}+u_d,
s_{l_{id}}^2
\right),
\]
with $b_{01}=1.84$, $b_{02}=2.61$, $s_1=0.74$, and $s_2=0.37$.

Repeated samples are then drawn from the synthetic finite population using a
domain-preserving stratified sampling design. Within
each domain \(d\), a fixed sampling fraction \(f=0.1\) is applied, yielding a domain sample size
\(n_d=\lceil fN_d\rceil\). The sample size \(n_d\) is then allocated
proportionally across the strata defined by the categorical covariate $x$.
Since proportional allocations are generally non-integer, integer stratum
sample sizes are obtained by taking the floor of the proportional allocation
and assigning the remaining units to the strata with the largest decimal
remainders. Finally,
simple random sampling without replacement is carried out independently within
each domain-by-\(x\) stratum. The resulting inclusion probability for unit \(i\)
in domain \(d\) and stratum \(h\) is
\(\pi_{idh}=n_{dh}/N_{dh}\), and the corresponding design weight is
\(w_{idh}=1/\pi_{idh}\).

After drawing the $m$-th sample $\mathbf{y}^{(m)}_s$, $m=1,\dots,250$, the corresponding coarsened response vector ${\mathbf{y}^{\star}}^{(m)}_s$ is generated by simulating the coarsening mechanism. Specifically, we use the proportional odds model introduced in \eqref{eq:probsG} and define two scenarios:
\begin{itemize}
    \item[] \textit{Scenario 1.} A non-ignorable coarsening mechanism: $\gamma_{01}=-10$, $\gamma_{02}=-7$, and $\gamma_1=-3.5$.
    \item[] \textit{Scenario 2.} An ignorable coarsening mechanism: $\gamma_{01}=-2$, $\gamma_{02}=0$, and $\gamma_1=0$.
\end{itemize}

As target quantities, we focus on the population summaries considered in Section \ref{sec:smokers}. To introduce a unified notation, let 
$\theta\in\{\pi^{\text{HS}}, \mu, Q^{0.5}, Q^{0.75}\}$ 
denote a generic target quantity at the overall population level, and let $\theta_d$ denote the corresponding quantity at the sub-population level. Since the finite population is fully known in the simulation setting, these quantities are available and can be used to assess the performance of the competing estimators.
Specifically, we compare the following estimation strategies:
\begin{itemize}
    \item The direct estimator applied to the fully observed sample $\mathbf{y}^{(m)}_s$. This strategy, labelled as \textit{Oracle}, is used as a benchmark, as it relies on information not available in practice.
    
    \item The \textit{Naive} direct estimator, namely the design-based estimator applied directly to the coarsened samples ${\mathbf{y}^{\star}}^{(m)}_s$.
    
    \item The CA estimator based on the simple Lognormal model (CA-LN), included to assess the robustness of the proposed methodology under model misspecification.
    
    \item The CA estimator based on the Lognormal mixture model (CA-LNM).
\end{itemize}

For each target quantity, the performance of the competing estimators is evaluated over the $M=250$ MC replications. Let 
$
\widehat{\theta}^{(m)}_{\text{Met}},
$
where 
$
\text{Met}\in\{\text{Oracle},\text{Naive},\text{CA-LN},\text{CA-LNM}\},
$
denotes the estimate of the generic population quantity $\theta$ obtained in the $m$-th replication under method $\text{Met}$. Similarly, let 
\(
\boldsymbol{\Theta}^{(m)}_\text{Met}=\left[L^{\theta(m)}_{\text{Met}},U^{\theta(m)}_{\text{Met}}\right]
\)
be the corresponding 95\% confidence interval. We consider the relative bias, the root mean squared error, the empirical coverage probability and the average interval width as performance measured, being defined as
\[
\operatorname{RB}(\widehat{\theta}^{(m)}_{\text{Met}})
=
\frac{1}{M}\sum_{m=1}^{M}
\frac{\widehat{\theta}^{(m)}_{\text{Met}}-\theta}{\theta},\quad
\operatorname{RMSE}(\widehat{\theta}^{(m)}_{\text{Met}})
=
\sqrt{
\frac{1}{M}\sum_{m=1}^{M}
\left(
\widehat{\theta}^{(m)}_{\text{Met}}-\theta
\right)^2
},
\]
\[
\operatorname{Cov}(\boldsymbol{\Theta}^{(m)}_\text{Met})
=
\frac{1}{M}\sum_{m=1}^{M}
\boldsymbol{1}
\left\{
\theta \in 
\boldsymbol{\Theta}^{(m)}_\text{Met}
\right\},\quad
\operatorname{Width}(\boldsymbol{\Theta}^{(m)}_\text{Met})
=
\frac{1}{M}\sum_{m=1}^{M}
\left(
U^{\theta(m)}_{\text{Met}}-L^{\theta(m)}_{\text{Met}}
\right).
\]
The same definitions are used for the sub-population quantities $\theta_d$, replacing $\theta$ with $\theta_d$ and $\widehat{\theta}^{(m)}_{\text{Met}}$ with $\widehat{\theta}^{(m)}_{d,\text{Met}}$.

\begin{figure}
    \centering
    \includegraphics[width=1\linewidth]{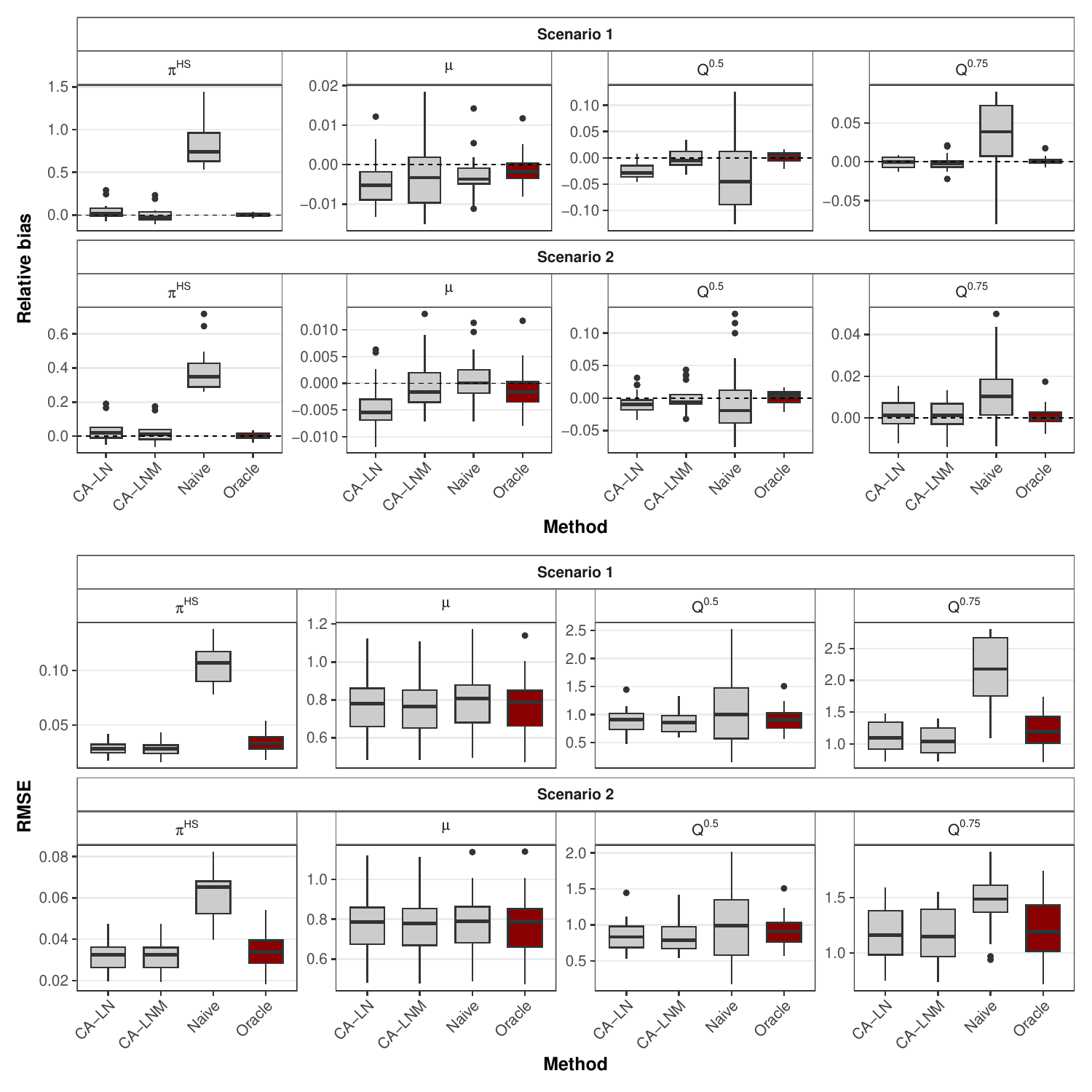}
    \caption{Relative bias and RMSE across simulation replicates, by coarsening scenario, target estimand and estimation method. The target estimands are $\pi^{HS}$, $\mu$, $Q^{0.5}$ and $Q^{0.75}$, while the methods compared are the oracle estimator, the naive direct estimator, and the LN- and LNM-based coarsening-adjusted estimators. The dashed horizontal line marks the absence of bias in 0.}
    \label{fig:placeholder_simbox}
\end{figure}

Figure~\ref{fig:placeholder_simbox} summarizes the distribution of relative bias and RMSE across sub-populations. The corresponding results for confidence intervals, in terms of frequentist coverage and average width, are reported in Figure~S5 in the Supplementary Material, while Table~S1 reports the simulation results for estimators of quantities concerning the whole population.

The most evident discrepancy concerns the HS proportion, \(\pi^{\text{HS}}\). In both scenarios, the naive estimator is severely biased upward, with relative bias equal to 0.739 under the non-ignorable coarsening mechanism and 0.353 under the ignorable mechanism. As a consequence, the corresponding confidence intervals fail to achieve nominal coverage. This result is consistent with the empirical evidence from the PASSI application: when a policy-relevant threshold coincides with a strong heaping point, treating reported values as exact leads to a substantial overestimation of the threshold-based indicator. The CA estimators substantially reduce this bias: both CA-LN and CA-LNM yield relative biases close to zero and coverage probabilities close to the nominal level. The CA-LNM estimator performs slightly better in terms of bias in Scenario~1, while the LN estimator remains competitive, suggesting some robustness to model misspecification.

A different pattern is observed for the mean. For this estimand, the naive, oracle, and CA estimators show very similar performance in both scenarios: relative biases are close to zero, RMSE values are comparable, and coverage probabilities remain close to the nominal level. This confirms that the mean is much less sensitive to heaping than threshold-based indicators.

Focusing on quantiles, the naive estimator is affected by the discreteness induced by heaping and, although its confidence intervals may sometimes display high coverage, this is achieved at the cost of substantially wider intervals, particularly in Scenario~1. The CA estimators, especially CA-LNM, reduce RMSE and produce interval widths closer to the oracle benchmark. This suggests that the proposed adjustment improves not only point estimation, but also uncertainty assessment for distributional summaries.

The comparison between the two coarsening scenarios further clarifies the role of the reporting mechanism. As expected, the consequences of ignoring coarsening are more severe under the non-ignorable mechanism, where the probability of coarser reporting depends on the latent value. However, the naive estimator remains problematic even under the ignorable scenario for \(\pi^{\text{HS}}\) and, to a lesser extent, for quantiles. This indicates that the issue is not only the dependence between the latent response and the reporting regime, but also the interaction between heaping points and the specific functional used for policy monitoring.

Overall, the simulation study supports the main message of the empirical application. Coarsening adjustment is essential for threshold-based indicators, such as the HS proportion, and for quantiles, while it is less consequential for means. Moreover, although the LNM model provides the best overall performance, as expected given the data-generating mechanism, the simpler LN model still performs satisfactorily for several targets. The proposed strategy, therefore, appears able to preserve the design-based target of inference while reducing bias and improving uncertainty assessment in the presence of coarsened self-reported responses.

\section{Discussion and concluding remarks}\label{sec:conclusion}

In this paper, we proposed a general procedure for estimating finite-population indicators from coarsened self-reported survey data under complex sampling designs. The observed response is treated as an incomplete manifestation of an underlying latent variable, and posterior predictive reconstruction is used to generate plausible reconstructed versions of the sampled data. Standard design-based estimators are then applied to these reconstructed samples. In this way, the inferential target remains the finite-population quantity that would be estimated if the latent responses were fully observed, while the uncertainty induced by coarsening is explicitly propagated.

A key feature of the proposed framework is the separation between sampling uncertainty, induced by the survey design, and reconstruction uncertainty, induced by the unobserved latent responses. This decomposition provides both a variance estimator and a diagnostic tool for assessing when coarsening matters for a given estimand. The fraction of uncertainty attributable to reconstruction can be viewed as close in spirit to the fraction of missing information in multiple imputation \citep{rubin1996multiple,murray2018multiple}, although here it refers to information loss due to coarse reporting rather than item nonresponse. This distinction is relevant because the additional uncertainty induced by reconstruction and the bias induced by ignoring coarsening need not move together.

The empirical and simulation results show that the consequences of coarsening are strongly estimand-specific. Means appear relatively robust, whereas threshold-based proportions are sensitive when heaping occurs at policy-relevant cut-points, and quantiles may inherit the artificial discreteness of the reported distribution. These findings suggest that coarsening should be evaluated with respect to the target finite-population functional, rather than through global diagnostics alone.

The proposed approach is broader than the specific applications considered here. It can be viewed as a reconstruction strategy for settings in which the observed data restrict the latent value to a set of admissible values, rather than revealing it exactly. This connects the framework to the wider literature on coarse data, which includes rounded, heaped, censored, partially categorized, and missing observations \citep{heitjan1991ignorability,heitjan1994ignorability}. A related, although different, setting is considered by
\citet{jenkins2011measuring}, who use multiple imputation to estimate
inequality measures from right-censored top-coded income data by reconstructing
censored observations before applying complete-data estimators.
%Truncated data are more delicate: when observations outside a given range are not observed at all, reconstruction must also model the mechanism determining inclusion in the observed sample. Thus, while censoring can often be handled by adapting the conditional reconstruction step, truncation requires assumptions on both the latent outcome distribution and the selection process.\\
% Io questo non starei a dirlo. Comunque l'idea della truncation sta dentro al coarsening, non starei quindi a distinguere

Several limitations should be acknowledged. First, the validity of the CA estimators depends on the adequacy of both the latent outcome model and the reporting model. Since the true latent values are not observed, these assumptions cannot be fully validated from the data; posterior predictive checks and sensitivity analyses are therefore important. Second, the pseudo-posterior implementation incorporates survey weights through a weighted likelihood, providing a practical way to account for informative sampling, but not a fully joint model for the sampling design and the response process. Third, although the mixture specification increases flexibility, the latent distribution is still modelled parametrically, so misspecification may remain an issue, especially in small domains or in the tails. Finally, inference for nonlinear functionals such as quantiles and threshold proportions requires care, as normal approximations may be less reliable near boundaries, sharp cut-points, or in domains with small effective sample sizes.

Future work could extend the framework in several directions. More flexible latent models could be considered, including semi-parametric mixtures, Bayesian nonparametric priors, or distributional regression models in which covariates affect both the location and the shape of the latent distribution \citep{murray2018multiple}. Further extensions could also focus on richer reporting models, allowing coarsening behaviour to vary over time, across interview modes, or across population groups. Finally, the approach could be generalized to multivariate coarsening, where several self-reported variables are rounded jointly or where derived indicators depend on multiple coarsened components, such as body mass index based on self-reported height and weight.

%Overall, this paper contributes a bridge between model-based reconstruction of incompletely observed responses and design-based inference for finite-population quantities. Its main goal is not to smooth heaped data for descriptive purposes, but to recover the estimators that would be used if the latent survey responses were available, while preserving the role of the sampling design and quantifying the additional uncertainty due to coarse observation. This makes the framework relevant beyond the motivating application, especially for survey settings in which official or scientific conclusions depend on nonlinear functionals of self-reported numerical variables.

\bibliographystyle{abbrvnat}
\bibliography{biblio.bib}

@article{heitjan1994ignorability,
  title={Ignorability in general incomplete-data models},
  author={Heitjan, Daniel F},
  journal={Biometrika},
  volume={81},
  number={4},
  pages={701-708},
  year={1994},
  publisher={Oxford University Press}
}

@article{jenkins2011measuring,
  title={Measuring inequality using censored data: a multiple-imputation approach to estimation and inference},
  author={Jenkins, Stephen P and Burkhauser, Richard V and Feng, Shuaizhang and Larrimore, Jeff},
  journal={Journal of the Royal Statistical Society Series A: Statistics in Society},
  volume={174},
  number={1},
  pages={63--81},
  year={2011},
  publisher={Oxford University Press}
}

@article{drechsler2016beat,
  title={Beat the heap: An imputation strategy for valid inferences from rounded income data},
  author={Drechsler, J{\"o}rg and Kiesl, Hans},
  journal={Journal of Survey Statistics and Methodology},
  volume={4},
  number={1},
  pages={22--42},
  year={2016},
  publisher={Oxford University Press}
}

@book{sarndal1992,
  author    = {S{\"a}rndal, Carl-Erik and Swensson, Bengt and Wretman, Jan},
  title     = {Model Assisted Survey Sampling},
  publisher = {Springer},
  year      = {1992}
}

@article{savitskytoth2016,
  author  = {Savitsky, Terrance D. and Toth, Daniell},
  title   = {Bayesian Estimation Under Informative Sampling},
  journal = {Electronic Journal of Statistics},
  year    = {2016},
  volume  = {10},
  number  = {1},
  pages   = {1677--1708}
}

@article{manski2010rounding,
  title={Rounding probabilistic expectations in surveys},
  author={Manski, Charles F and Molinari, Francesca},
  journal={Journal of Business \& Economic Statistics},
  volume={28},
  number={2},
  pages={219--231},
  year={2010},
  publisher={Taylor \& Francis}
}

@article{carpenter2017stan,
  title={Stan: A probabilistic programming language},
  author={Carpenter, Bob and Gelman, Andrew and Hoffman, Matthew D and Lee, Daniel and Goodrich, Ben and Betancourt, Michael and Brubaker, Marcus and Guo, Jiqiang and Li, Peter and Riddell, Allen},
  journal={Journal of statistical software},
  volume={76},
  pages={1--32},
  year={2017}
}

@article{woodruff1952confidence,
  title={Confidence intervals for medians and other position measures},
  author={Woodruff, Ralph S},
  journal={Journal of the American Statistical Association},
  volume={47},
  number={260},
  pages={635--646},
  year={1952},
  publisher={Taylor \& Francis}
}

@article{isaki1982survey,
  title={Survey design under the regression superpopulation model},
  author={Isaki, Cary T and Fuller, Wayne A},
  journal={Journal of the American Statistical Association},
  volume={77},
  number={377},
  pages={89--96},
  year={1982},
  publisher={Taylor \& Francis}
}

@article{gardini2026coarsened,
  title={Coarsened data in small area estimation: a Bayesian two-part model for mapping smoking behaviour},
  author={Gardini, Aldo and Mori, Lorenzo},
  journal={arXiv preprint arXiv:2601.19729},
  year={2026}
}

@online{eurostat_tobacco_2022,
  author  = {{Eurostat}},
  title   = {Tobacco consumption statistics},
  year    = {2022},
  url     = {https://ec.europa.eu/eurostat/statistics-explained/index.php?title=Tobacco_consumption_statistics},
  note    = {Statistics Explained. Data extracted in May 2022; data refer to 2019. Accessed: 20 May 2026}
}

@article{lumley2004survey,
  author  = {Thomas Lumley},
  title   = {Analysis of Complex Survey Samples},
  journal = {Journal of Statistical Software},
  year    = {2004},
  volume  = {9},
  number  = {1},
  pages   = {1--19}
}

@article{sitter2001note,
  title={A note on Woodruff confidence intervals for quantiles},
  author={Sitter, Randy R and Wu, Changbao},
  journal={Statistics \& probability letters},
  volume={52},
  number={4},
  pages={353--358},
  year={2001},
  publisher={Elsevier}
}

@article{heitjan1991ignorability,
  title={Ignorability and coarse data},
  author={Heitjan, Daniel F and Rubin, Donald B},
  journal={The annals of statistics},
  pages={2244--2253},
  year={1991},
  publisher={JSTOR}
}

@article{wang2008modeling,
  title={Modeling heaping in self-reported cigarette counts},
  author={Wang, Hao and Heitjan, Daniel F},
  journal={Statistics in medicine},
  volume={27},
  number={19},
  pages={3789--3804},
  year={2008},
  publisher={Wiley Online Library}
}

@article{zinn2016heaping,
  author = {Zinn, S. and Würbach, A.},
  title = {A statistical approach to address the problem of heaping in self-reported income data},
  journal = {Journal of Applied Statistics},
  volume = {43},
  number = {4},
  pages = {682--703},
  year = {2016}
}

@misc{iss_2023_protocollo_passi,
  author       = {{Istituto Superiore di Sanità}},
  title        = {Protocollo di studio PASSI (Versione 8 febbraio 2023)},
  year         = {2023},
  month        = feb,
  howpublished = {\url{https://www.epicentro.iss.it/passi/infoPassi/protocollo-operativo-passi}},
  note         = {Accessed: 2025-09-09}
}

@misc{un2014fundamental,
  author = {{United Nations}},
  title  = {Fundamental Principles of Official Statistics},
  year   = {2014},
  note   = {United Nations General Assembly Resolution 68/261}
}

@article{heitjan1990inference,
  title={Inference from coarse data via multiple imputation with application to age heaping},
  author={Heitjan, Daniel F and Rubin, Donald B},
  journal={Journal of the American Statistical Association},
  volume={85},
  number={410},
  pages={304--314},
  year={1990},
  publisher={Taylor \& Francis}
}

@book{remington2010chronic,
  editor    = {Remington, Patrick L. and Brownson, Ross C. and Wegner, Mary V.},
  title     = {Chronic Disease Epidemiology, Prevention, and Control},
  edition   = {3},
  publisher = {American Public Health Association},
  year      = {2010}
}

@misc{cdc_nhis,
  author = {{National Center for Health Statistics}},
  title  = {National Health Interview Survey},
  year   = {2025},
  note   = {Centers for Disease Control and Prevention}
}

@book{tourangeau2000psychology,
  author    = {Tourangeau, Roger and Rips, Lance J. and Rasinski, Kenneth},
  title     = {The Psychology of Survey Response},
  publisher = {Cambridge University Press},
  year      = {2000}
}

@incollection{bound2001measurement,
  author    = {Bound, John and Brown, Charles and Mathiowetz, Nancy},
  title     = {Measurement Error in Survey Data},
  booktitle = {Handbook of Econometrics},
  volume    = {5},
  editor    = {Heckman, James J. and Leamer, Edward},
  publisher = {Elsevier},
  pages     = {3705--3843},
  year      = {2001}
}

@article{gross2016kernel,
  author  = {Groß, Marcus and Rendtel, Ulrich},
  title   = {Kernel Density Estimation for Heaped Data},
  journal = {Journal of Survey Statistics and Methodology},
  volume  = {4},
  number  = {3},
  pages   = {339--361},
  year    = {2016}
}

@misc{iss_passi_temi,
  author = {{Istituto Superiore di Sanit{\`a}}},
  title  = {Gli ambiti della sorveglianza PASSI},
  year   = {2025},
  note   = {EpiCentro},
  url    = {https://www.epicentro.iss.it/passi/dati/temi}
}

@misc{eurostat_ehis,
  author = {{Eurostat}},
  title  = {European Health Interview Survey (EHIS) -- Microdata},
  year   = {2025},
  note   = {European Commission, Eurostat},
  url    = {https://ec.europa.eu/eurostat/web/microdata/collections-research/european-health-interview-survey}
}

@book{who2020physicalactivity,
  author    = {{World Health Organization}},
  title     = {WHO guidelines on physical activity and sedentary behaviour},
  year      = {2020},
  publisher = {World Health Organization},
  address   = {Geneva},
  isbn      = {9789240015128},
  url       = {https://www.who.int/publications/i/item/9789240015128},
  note      = {Published 25 November 2020}
}

@article{rubin1996multiple,
  title={Multiple imputation after 18+ years},
  author={Rubin, Donald B},
  journal={Journal of the American statistical Association},
  volume={91},
  number={434},
  pages={473--489},
  year={1996},
  publisher={Taylor \& Francis}
}

@article{murray2018multiple,
  title={Multiple Imputation: A Review of Practical and Theoretical Findings},
  author={Murray, Jared S},
  journal={Statistical Science},
  volume={33},
  number={2},
  year={2018}
}

\end{document}